\documentclass[aps,prl,twocolumn,nopacs,superscriptaddress]{revtex4}
\usepackage[utf8]{inputenc}

\usepackage{graphicx}  
\usepackage{dcolumn}   
\usepackage{bm}        
\usepackage{amssymb}   
\usepackage{amsmath}
\usepackage{units}
\usepackage[english]{babel}
\usepackage[dvipsnames]{xcolor}
\usepackage{xfrac}

\usepackage[%
colorlinks=true,
urlcolor=RoyalBlue,
linkcolor=RoyalBlue,
citecolor=RoyalBlue,
]{hyperref}

\usepackage{sidecap}
\sidecaptionvpos{figure}{m}  

\usepackage[type1]{libertine}                                        
\usepackage{textcomp}
\usepackage[scaled=.85]{beramono}
\usepackage[libertine,cmintegrals,cmbraces,vvarbb,slantedGreek]{newtxmath}
\usepackage[scr=boondoxo]{mathalfa}
\usepackage{bm}
\usepackage[lf]{carlito}

\usepackage{braket}
\usepackage{bm}

\makeatletter
\DeclareRobustCommand{\cev}[1]{%
  \mathpalette\do@cev{#1}%
}
\newcommand{\do@cev}[2]{%
  \fix@cev{#1}{+}%
  \reflectbox{$\m@th#1\vec{\reflectbox{$\fix@cev{#1}{-}\m@th#1#2\fix@cev{#1}{+}$}}$}%
  \fix@cev{#1}{-}%
}
\newcommand{\fix@cev}[2]{%
  \ifx#1\displaystyle
    \mkern#23mu
  \else
    \ifx#1\textstyle
      \mkern#23mu
    \else
      \ifx#1\scriptstyle
        \mkern#22mu
      \else
        \mkern#22mu
      \fi
    \fi
  \fi
}
\makeatother

\newcommand{\spinhalf}{spin-\sfrac{1}{2}}

\newcommand{\panel}[1]{(#1)}
\newcommand{\panelcaption}[1]{(#1)}
\newcommand{\panelsubcaption}[1]{(#1)}

\begin{document}

\title{Superconducting Quantum Interference at the Atomic Scale}

\author{Sujoy Karan}
\affiliation{Max-Planck-Institut f\"ur Festk\"orperforschung, Heisenbergstraße 1,
70569 Stuttgart, Germany}
\author{Haonan Huang}
\affiliation{Max-Planck-Institut f\"ur Festk\"orperforschung, Heisenbergstraße 1,
70569 Stuttgart, Germany}
\author{Ciprian Padurariu}
\affiliation{Institut für Komplexe Quantensysteme and IQST, Universität Ulm, Albert-Einstein-Allee 11, 89069 Ulm, Germany}
\author{Bj\"orn Kubala}
\affiliation{Institut für Komplexe Quantensysteme and IQST, Universität Ulm, Albert-Einstein-Allee 11, 89069 Ulm, Germany}
\affiliation{Institute of Quantum Technologies, German Aerospace Center (DLR), S\"oflinger Stra{\ss}e 100, 89077 Ulm, Germany}
\author{Andreas Theiler}
\affiliation{Department of Physics and Astronomy, Uppsala University, Box 516, 751\,20 Uppsala, Sweden}
\author{Annica M. Black-Schaffer}
\affiliation{Department of Physics and Astronomy, Uppsala University, Box 516, 751\,20 Uppsala, Sweden}
\author{Gonzalo Morr\'as}
\affiliation{Departamento de F\'{\i}sica Te\'orica de la Materia Condensada and
Condensed Matter Physics Center (IFIMAC), Universidad Aut\'onoma de Madrid, 28049 Madrid, Spain}
\author{Alfredo Levy Yeyati}
\affiliation{Departamento de F\'{\i}sica Te\'orica de la Materia Condensada and
Condensed Matter Physics Center (IFIMAC), Universidad Aut\'onoma de Madrid, 28049 Madrid, Spain}
\author{Juan Carlos Cuevas}
\affiliation{Departamento de F\'{\i}sica Te\'orica de la Materia Condensada and
Condensed Matter Physics Center (IFIMAC), Universidad Aut\'onoma de Madrid, 28049 Madrid, Spain}
\author{Joachim Ankerhold}
\affiliation{Institut für Komplexe Quantensysteme and IQST, Universität Ulm, Albert-Einstein-Allee 11, 89069 Ulm, Germany}
\author{Klaus Kern}
\affiliation{Max-Planck-Institut f\"ur Festk\"orperforschung, Heisenbergstraße 1,
70569 Stuttgart, Germany}
\affiliation{Institut de Physique, Ecole Polytechnique Fédérale de Lausanne, 1015 Lausanne, Switzerland}
\author{Christian R. Ast}
\email[Corresponding author; electronic address:\ ]{c.ast@fkf.mpg.de}
\affiliation{Max-Planck-Institut f\"ur Festk\"orperforschung, Heisenbergstraße 1,
70569 Stuttgart, Germany}

\date{\today}

\begin{abstract}
A single spin in a Josephson junction can reverse the flow of the supercurrent. At mesoscopic length scales, such $\pi$-junctions are employed in various instances from finding the pairing symmetry to quantum computing. In Yu-Shiba-Rusinov (YSR) states, the atomic scale counterpart of a single spin in a superconducting tunnel junction, the supercurrent reversal so far has remained elusive. Using scanning tunneling microscopy (STM), we demonstrate such a 0 to $\pi$ transition of a Josephson junction through a YSR state as we continuously change the impurity-superconductor coupling. We detect the sign change in the critical current by exploiting a second transport channel as reference in analogy to a superconducting quantum interference device (SQUID), which provides the STM with the required phase sensitivity. The measured change in the Josephson current is a signature of the quantum phase transition and allows its characterization with unprecedented resolution.
\end{abstract}

\maketitle

Two superconductors that are connected by a weak link can sustain a supercurrent, which is carried by Cooper pairs --- the well-known Josephson effect \cite{josephson_possible_1962}. Inserting a single spin into the junction may completely change its behavior by reversing the direction of the supercurrent \cite{kulik_magnitude_1966}, which is the result of a $\pi$-shift in the phase across the junction. Such $\pi$-junctions have been used in finding the pairing symmetry in unconventional superconductors \cite{tsuei_pairing_1994,kirtley_symmetry_1995,harlingen_phase-sensitive_1995,wollman_evidence_1995,tsuei_pairing_2000} and they have been proposed as building blocks for energy efficient quantum computing or high-speed memory \cite{gingrich_controllable_2016,cleuziou_carbon_2006,feofanov_implementation_2010}. At mesoscopic length scales ($\approx$10 to 100\,nm), $\pi$-junctions may be realized by singly occupied quantum dots or ferromagnetic interlayers \cite{ryazanov_coupling_2001,kontos_josephson_2002,robinson_critical_2006,shimizu_multilevel_1998,dam_supercurrent_2006,saldana_two-impurity_2020,martin-rodero_josephson_2011,franceschi_hybrid_2010}. At the atomic scale ($\approx$0.1\,nm), a single magnetic impurity, which is exchange coupled to a superconductor, induces a  spin nondegenerate superconducting bound state, a Yu-Shiba-Rusinov (YSR) state \cite{yu_bound_1965,shiba_classical_1968,rusinov_superconductivity_1969}. By tuning the magnetic exchange coupling, the YSR state can be driven through a quantum phase transition (QPT) with a concomitant $\pi$-shift \cite{salkola_spectral_1997,flatte_local_1997,flatte_local_1997-1}.

The hallmark of this quantum phase transition (QPT) in YSR states is a discontinuous change in the total spin of the respective ground states: a previously free impurity spin turns into a screened spin, when the magnetic exchange coupling increases beyond a critical value. Consequently, a reversal in the flow of Cooper pairs through a YSR state has been predicted \cite{balatsky_impurity-induced_2006}. Experimentally, the QPT can be identified by a zero energy crossing of the YSR state in differential conductance spectra \cite{farinacci_tuning_2018,malavolti_tunable_2018,franke_competition_2011,bauer_microscopic_2013,kamlapure_investigation_2019}. However, the actual consequences for the fundamental Josephson effect remain elusive in atomic scale junctions.

The observation of a YSR state based $\pi$-junction is experimentally challenging, because detecting such a phase shift between superconducting ground states requires a reference channel. At mesoscopic length scales, this is typically solved by employing a superconducting quantum interference device (SQUID) loop geometry \cite{shimizu_multilevel_1998,dam_supercurrent_2006,saldana_two-impurity_2020,martin-rodero_josephson_2011}. To reach similar conditions at the atomic scale, a scanning tunneling microscope (STM) requires a rudimentary phase sensitivity through an additional transport channel \cite{dam_supercurrent_2006,franceschi_hybrid_2010}.

\begin{SCfigure*}
    \centering
    \includegraphics[width=0.75\textwidth]{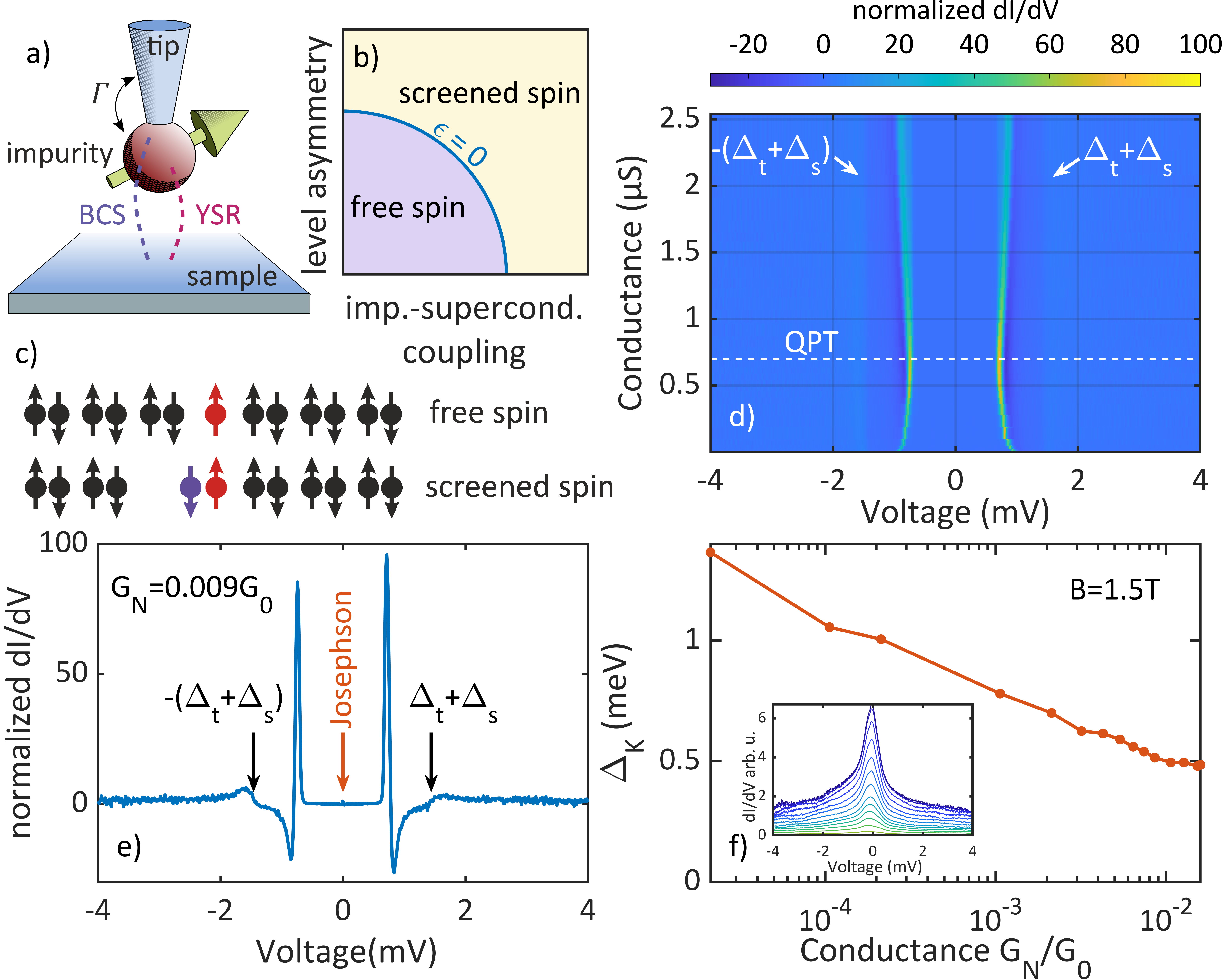}
    \caption{\textbf{Atomic YSR state.} \panelcaption{a} Schematic of the tunnel junction. The YSR impurity is at the tip with two transport channels (BCS and YSR) indicated as dashed lines. \panelcaption{b} Phase diagram of the YSR system as function of impurity-superconductor coupling and level (particle-hole) asymmetry. \panelcaption{c} Schematic of the free spin and the screened spin regime. In the screened regime, a Cooper pair is broken changing the overall parity of the system. \panelcaption{d} Differential conductance spectra as function of bias voltage ($x$-axis) and conductance ($y$-axis). The prominent peaks are the YSR states, while the coherence peaks are only barely visible. \panelcaption{e} Horizontal line cut through \panelsubcaption{d} to show the prominent YSR peaks. \panelcaption{f} Half-width at half maximum $\Delta_\text{K}$ of the Kondo peak in the same junction at a magnetic field of 1.5\,T, when superconductivity is quenched. The Kondo spectra are shown in the inset. }
    \label{fig:intro}
\end{SCfigure*}

Here, we demonstrate a supercurrent reversal in an atomic scale Josephson junction through a YSR state as we move across the QPT. We produce a magnetic impurity at the apex of a superconducting vanadium tip (see Fig.\ \ref{fig:intro}\panel{a}), which is approached to a superconducting V(100) sample. As we approach, the atomic forces pull on the impurity \cite{huang_quantum_2020,farinacci_tuning_2018,kezilebieke_observation_2019,ternes_interplay_2011}, which reduces the impurity-superconductor coupling $\Gamma$ along with the magnetic exchange coupling. This concomitantly allows the YSR state to pass from the strong scattering (screened spin) to the weak scattering (free spin) regime as outlined in Fig.\ \ref{fig:intro}\panel{b}. The two scenarios are schematically illustrated in Fig.\ \ref{fig:intro}\panel{c}, where the total spin in the free spin regime is $S_\text{tot}=\sfrac{1}{2}$. In the screened spin regime, a Cooper pair is broken to screen the impurity spin changing the overall parity of the system (indicating whether the total number of particles is even or odd) as well as the total spin to $S_\text{tot}=0$.

To detect the supercurrent reversal, we exploit the parallel presence of a second transport channel featuring a conventional superconducting Bardeen-Cooper-Schrieffer (BCS) gap without any YSR state as a reference channel (see Fig.\ \ref{fig:intro}\panel{a}). The sign change in the supercurrent through the YSR state manifests itself as a step in the measured net Josephson current resulting from the changeover of a constructive to a destructive interference of the two transport channels across the QPT.

The evolution of the YSR state as a function of the normal state conductance $G_\text{N}$ is shown in Fig.\ \ref{fig:intro}\panel{d}. The YSR state moves across the QPT when the YSR energies are closest to each other. Because both tip and sample are superconducting, in the spectrum the tip YSR states appear at voltages $V$ shifted by the sample gap $\Delta_\text{s}$, i.e.\ $eV=\varepsilon + \Delta_\text{s}$ with the YSR state energy $\varepsilon$ varying with the normal state conductance $G_\text{N}$. Interestingly, there are no distinct coherence peaks visible at the sum of the tip gap and the sample gap $\pm(\Delta_\text{t}+\Delta_\text{s})$, which indicates that a second transport channel through an empty gap (i.e.\ without any YSR state and hence with coherence peaks) has a much weaker, but still finite transmission compared to the YSR state. This can be more directly seen in a single spectrum near the QPT, which is shown in Fig.\ \ref{fig:intro}\panel{e}. The coherence peaks at $eV = \Delta_\text{t}+\Delta_\text{s}$ are greatly reduced and the YSR peaks are prominently enhanced by almost a factor of 100.

To confirm that the impurity-superconductor coupling (in our case the impurity-tip coupling) decreases with increasing conductance, we have measured the Kondo effect in the same junction by quenching the superconductivity in a magnetic field of 1.5\,T. The Kondo spectra are shown in the inset of Fig.\ \ref{fig:intro}\panel{f}, from which we extract the half width at half maximum $\Delta_\text{K}$. As $\Delta_\text{K}$ is directly related to the Kondo temperature and the magnetic exchange coupling, we conclude that the impurity-superconductor coupling decreases with decreasing tip-sample distance (i.e.\ increasing junction conductance). Physically, the impurity is pulled away from the tip by the attractive atomic forces of the approaching sample substrate \cite{ternes_interplay_2011}.

\begin{SCfigure*}
    \centering
    \includegraphics[width=0.65\textwidth]{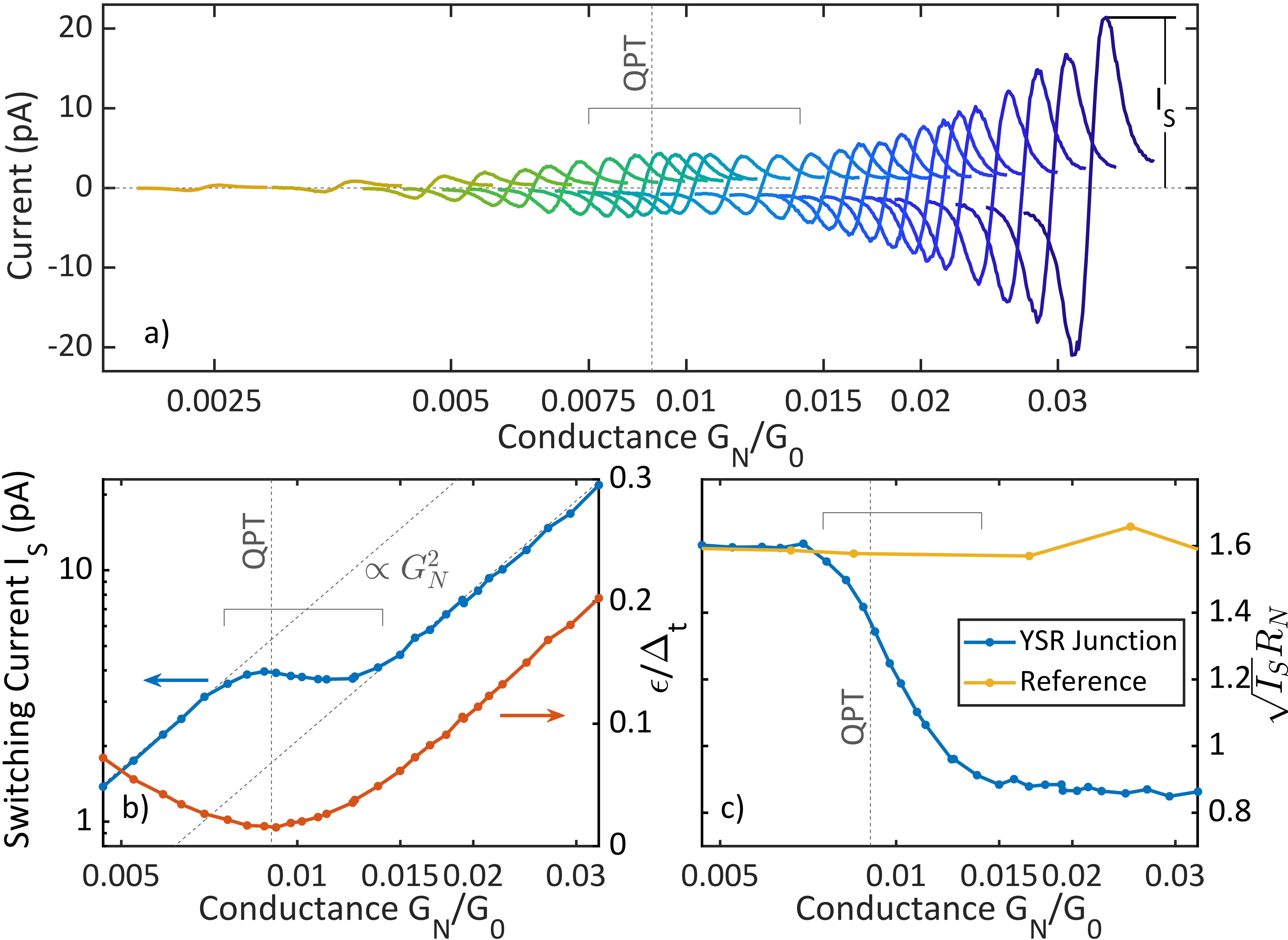}
    \caption{\textbf{Josephson Effect.} \panelcaption{a} Josephson spectra $I(V)$ in a range of $\pm 60\,\upmu$eV shifted horizontally by the conductance at which they were measured. The horizontal bracket indicates the region, where the evolution deviates from the conventional Ambegaokar-Baratoff formula. The quantum phase transition (QPT) is indicated by the vertical dashed line. \panelcaption{b} The switching current $I_\text{S}$, which is the current maximum indicated in \panelsubcaption{a} as function of the normal state conductance $G_\text{N}$ (blue line). The square dependence at low and high conductance is indicated by dashed lines (labeled by $\propto G_\text{N}^2$). The YSR energy as function of conductance is shown as a blue line. The minimum indicates the QPT (vertical dashed line). \panelcaption{c} This graph shows $\sqrt{I_\text{S}}R_\text{N}$ (blue line) of the data in \panelsubcaption{b}, which is proportional to $I_\text{C}R_\text{N}$ for a harmonic energy-phase relation ($R_\text{N}$: renormalized resistance (see text), $I_\text{C}$: critical current). A reference junction without any YSR states is shown as a yellow line indicating the expected evolution according to the Ambegaokar-Baratoff formula.}
    \label{fig:data}
\end{SCfigure*}

The supercurrent, which is carried by tunneling Cooper pairs (Josephson effect), is visible throughout the range of conductance values (see red arrow in Fig.\ \ref{fig:intro}\panel{e}). In the dynamical Coulomb blockade (DCB) regime, in which the STM operates \cite{ast_sensing_2016}, the typical voltage-biased measurement shows a negative current peak followed by a positive current peak of equal size near zero bias voltage. The evolution of the Josephson effect as function of conductance is shown in Fig.\ \ref{fig:data}\panel{a}. Each spectrum is shown in a bias voltage range of $\pm 60\,\mu$eV and offset horizontally. Assuming a harmonic current-phase relation in the DCB (i.e.\ $I(\varphi)=I_\text{C}\sin\varphi$, where $I_\text{C}$ is the critical current), the Josephson current is predicted to scale with the square of the critical current, i.e.\ $I(V)\propto I_\text{C}^2 \propto G_\text{N}^2$ \cite{senkpiel_single_2020,ingold_cooper-pair_1994,devoret_effect_1990,averin_incoherent_1990}. It can be directly seen in the data that this square dependence is not fulfilled in the data set in Fig.\ \ref{fig:data}\panel{a}. In particular, the region indicated by the horizontal bracket shows significant deviations, even a slight decrease in the Josephson current with increasing conductance. The conductance at which the QPT occurs is indicated by a vertical dashed line, which falls directly into the region of the horizontal bracket.

For a more quantitative analysis, we plot the current maxima $I_\text{S}$ (switching current) for each conductance as a blue line in a double logarithmic plot in Fig.\ \ref{fig:data}\panel{b}. The expected square dependence on the conductance ($I_\text{S}\propto I_\text{C}^2 \propto G_\text{N}^2$) can be clearly seen for very small and very large conductances. In the transition region (indicated by the horizontal bracket), the behavior of the switching current $I_\text{S}$ strongly changes. For comparison, we plot the experimentally extracted energies of the YSR state (red line), which has a minimum at the QPT (vertical dashed line) \cite{gapnote}. This indicates a drastic change in the behavior of the Josephson effect across the QPT.

To put the evolution of the switching current in reference to other Josephson junctions, we calculate $\sqrt{I_\text{S}}R_\text{N}$, which is shown in Fig.\ \ref{fig:data}\panel{c} ($R_\text{N}$ is the normal state tunneling resistance). This quantity is proportional to the product $I_\text{C}R_\text{N}$ for a harmonic energy-phase relation. In this way, the overall conductance dependence is eliminated such that the measurement appears like a step in Fig.\ \ref{fig:data}\panel{c} with a sizeable reduction in height almost by a factor of two across the QPT. We will show below that this is due to a supercurrent reversal in the YSR channel which leads to a crossover from a constructive to a destructive interference between the two transport channels.

In order to compare the experimental data to the theory, we have to renormalize the normal state resistance $R_\text{N}$ for the YSR spectra due to the enhanced density of states from Kondo correlations (for details see the Supplementary Information \cite{supinf}). The reference spectra (orange line in Fig.\ \ref{fig:data}\panel{c}) are measured for a Josephson junction without any YSR states, where the $\sqrt{I_\text{S}}R_\text{N}$ is constant as expected from the Ambegaokar-Baratoff formula \cite{jack_critical_2016,ambegaokar_tunneling_1963}.

\begin{SCfigure*}
    \centering
    \includegraphics[width=0.55\textwidth]{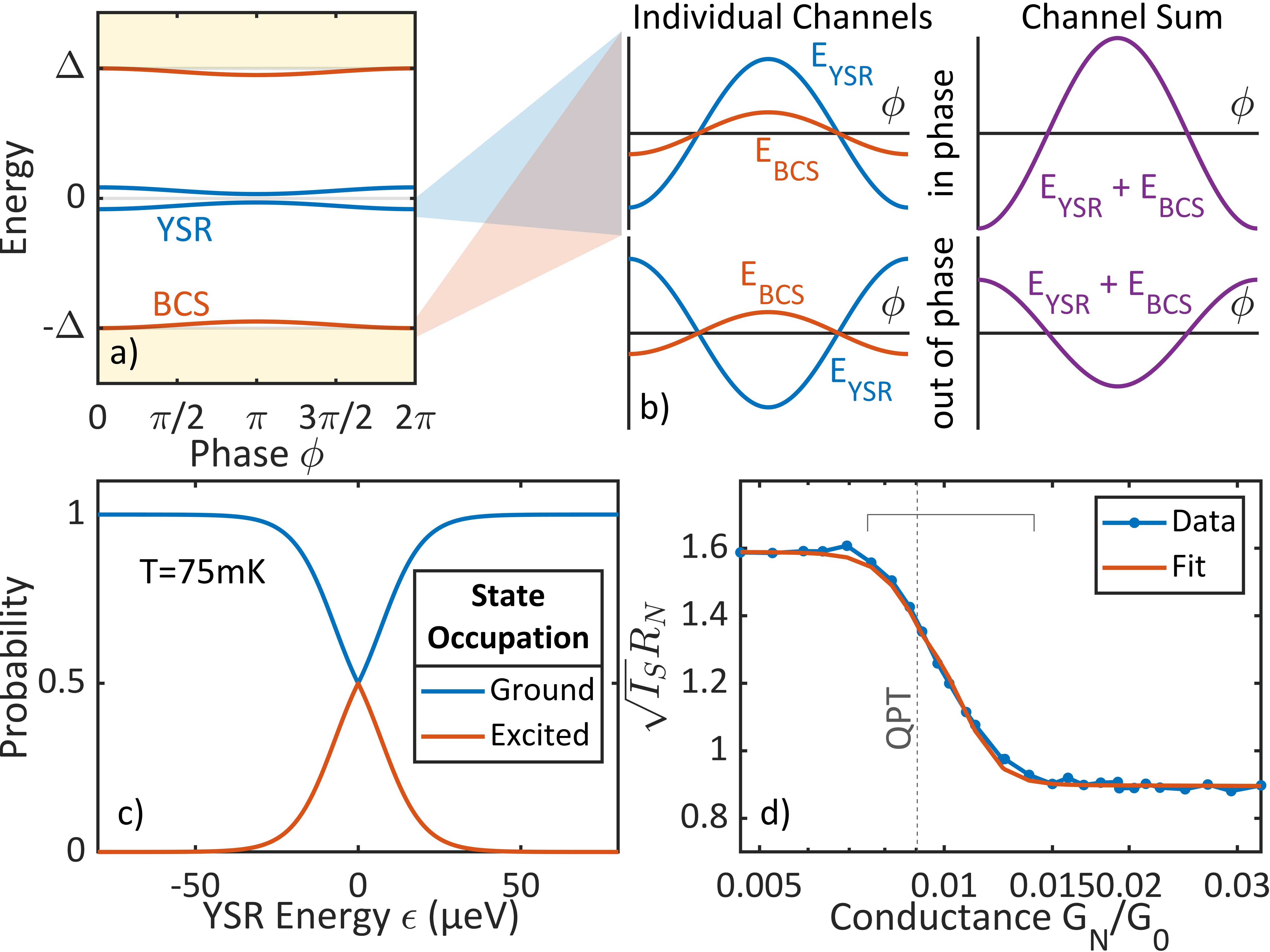}
    \caption{\textbf{Supercurrent Reversal.} \panelcaption{a} Energy-phase relation for the BCS channel (red) and the YSR channel (blue). For the Josephson current only the oscillation is relevant, but not the energy offset. \panelcaption{b} shows a zoom-in to the oscillation for the two channels. The upper two panels in \panelsubcaption{b} show the in-phase oscillation in the screened spin regime. The lower two panels in \panelsubcaption{b} show the out-of-phase oscillation in the free spin regime, which is indicative of the supercurrent reversal. The STM is not sensitive of the sign of the supercurrent, but the concomitant change in magnitude is clearly observable. \panelcaption{c} Probabilities for the system to be in the ground state (blue) or the excited state (red) as function of YSR state energy. \panelcaption{d} The $\sqrt{I_\text{S}}R_\text{N}$ product comparing experimental data with a theoretical fit. The only free parameters are the temperature, which defines the width of the QPT and the relative channel transmission, which defines the step height.}
    \label{fig:theory}
\end{SCfigure*}

To understand the behavior of the Josephson effect in Fig.\ \ref{fig:data}, we first have a look at the energy-phase relations far away from the QPT at high and at low conductance. In Fig.\ \ref{fig:theory}\panel{a}, the energy-phase relations for the BCS channel and the YSR channel, which is calculated from a mean field Anderson impurity model, are shown in red and blue, respectively (for details see the Supplementary Information \cite{supinf}). To calculate the energy-phase relation, we apply a constant phase difference $\varphi$ across the tunnel junction, but no bias voltage. A Fourier expansion of the energy-phase relation reveals that the most relevant contribution to the Josephson effect is the harmonic term proportional to $\cos(\varphi)$. Zooming in to both channels (cf.\ Fig.\ \ref{fig:theory}\panel{b}), we estimate that the ratio of the channel transmissions is about 4:1 (YSR:BCS): This results in a significantly smaller amplitude for the energy-phase relation of the BCS channel (red) than in the YSR channel (blue) (Individual Channels in Fig.\ \ref{fig:theory}\panel{b}). The coherent superposition of these two channels (Channel Sum in Fig.\ \ref{fig:theory}\panel{b}) leads to an overall sign change as well as different amplitudes, when the channels are in phase (upper row in Fig.\ \ref{fig:theory}\panel{b}) or out of phase (lower row in Fig.\ \ref{fig:theory}\panel{b}). In the measurement, we are only sensitive to the change in amplitude $I(V) \propto (E_\text{YSR}+E_\text{BCS})^2$ though, which results in the obvious step in Fig.\ \ref{fig:data}\panel{c}. We attribute the width of the step to the finite temperature in our experiment.

Since the temperature in our experiment (10\,mK) is still non-zero, we expect fluctuations due to thermal excitations close to the QPT. The probability for the system to be in the ground state (blue) or the excited state (orange) is indicated in Fig.\ \ref{fig:theory}\panel{c} using an effective temperature of 75\,mK. This will broaden the expected sharp features associated with the quantum phase transition. Taking the excitation probability due to the finite temperature into account, we can calculate the expected Josephson current in the DCB regime (see Supplementary Information \cite{supinf}). The fit is shown in Fig.\ \ref{fig:theory}\panel{d} with excellent agreement to the data. The only free parameters are the effective temperature $T_\text{eff}=75\,$mK, which is determined by the width of the transition and the ratio of the two channel transmissions, which is determined by the step height. For a best fit, we find that the YSR channel contributes 78.4\% and the BCS reference channel contributes 21.6\% to the total conductance relevant to the Josephson effect, which is consistent with the prominent YSR states and the strongly reduced coherence peaks in the quasiparticle spectra (see Fig.\ \ref{fig:intro}\panel{e}). All other parameters are given by the experimentally extracted values. In this way, we demonstrate that the supercurrent through an atomic scale YSR state reverses upon crossing the QPT, which can be detected in the STM by means of a BCS reference channel in analogy to a SQUID geometry (see Supplementary Information for more details \cite{supinf}).

\begin{figure}
    \centering
    \includegraphics[width=\columnwidth]{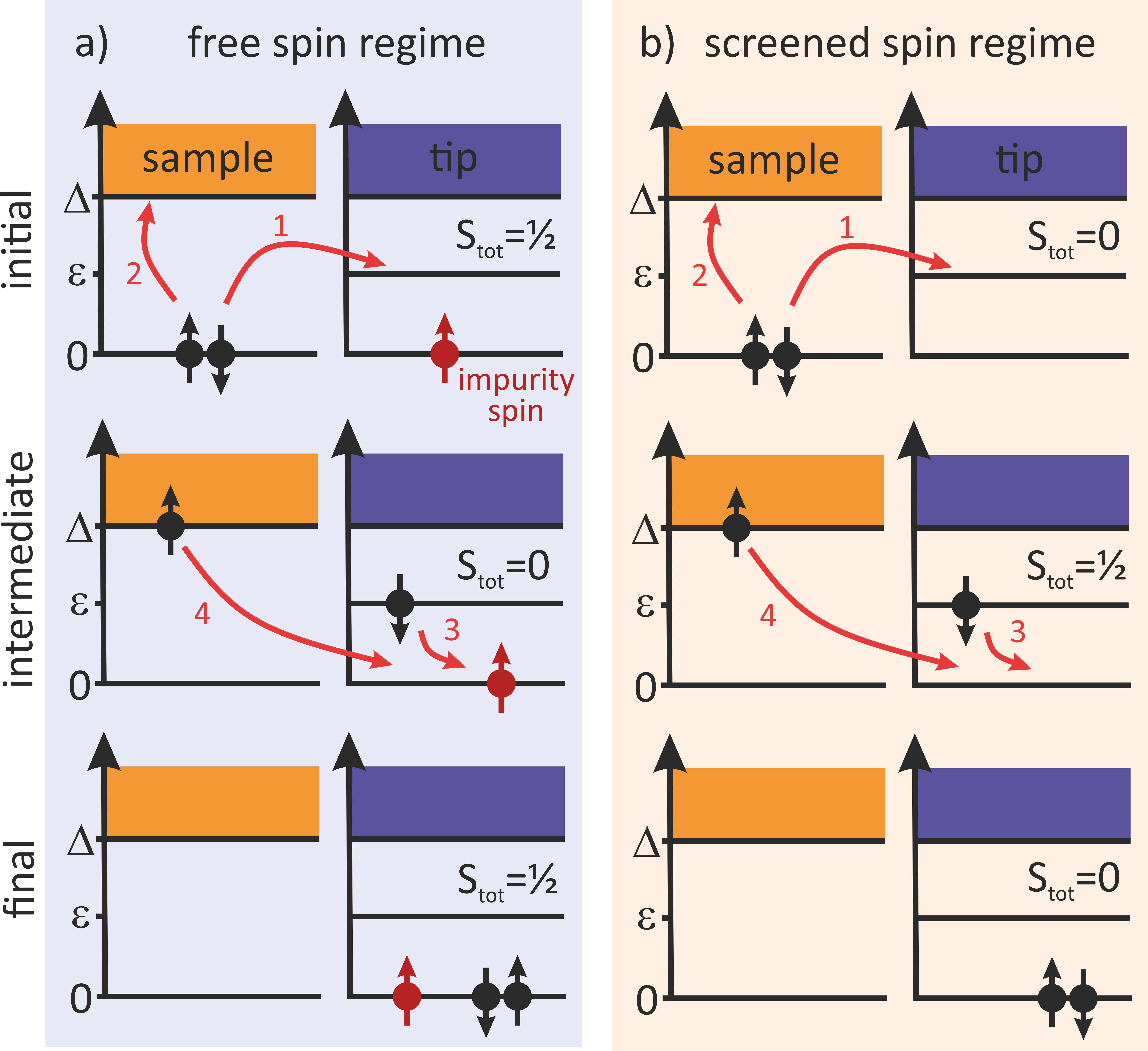}
    \caption{\textbf{Cooper Pair Tunneling.} The tunneling process in the free spin and in the screened spin regime is shown from the initial to the final state via an intermediate (virtual) state in the excitation picture. The order is set by the numbered red arrows. The total spin $S_\text{tot}$ describes the total spin of the YSR system including the spin of the impurity. \panelcaption{a} In the free spin regime, the order of the spins is exchanged compared to the intial state, which results in the supercurrent reversal. \panelcaption{b} In the screened spin regime, the order of the spins is retained, such that there is no sign change in the supercurrent.}
    \label{fig:schematic}
\end{figure}

To better understand the origin of this supercurrent reversal and to illustrate the crucial role of the impurity spin, we discuss the Cooper pair tunneling process in Fig.\ \ref{fig:schematic} using the excitation picture.  Zero energy denotes the ground state, $\varepsilon$ is the energy of the excited YSR state, and $\Delta$ marks the beginning of the quasiparticle continuum. The order is given by the numbered red arrows. Figure \ref{fig:schematic}\panel{a} describes the free spin regime, where the total spin is $S_\text{tot}=\sfrac{1}{2}$ \cite{balatsky_impurity-induced_2006,salkola_spectral_1997}. This indicates that the total parity (superconductor + impurity) must be odd. The Cooper pair transfer process involves a swap between two fermions, the one associated with the impurity and one associated with the Cooper pair, as depicted by the arrows 3 and 4 in Fig.\ \ref{fig:schematic}\panel{a}. Formally, this appears as an exchange of fermion operators inducing a negative sign ($\pi$-shift) \cite{dam_supercurrent_2006,spivak_negative_1991}. By contrast, Fig.\ \ref{fig:schematic}\panel{b} shows the screened spin regime, which has a ground state with total spin $S_\text{tot}=0$ and correspondingly even parity. Here, the Cooper pair transfers conventionally as in an empty BCS gap except that the YSR state is used as intermediate (virtual) state instead of the continuum. It is this switching between transport regimes when crossing the QPT, which we observe experimentally. Recalling Fig.\ \ref{fig:intro}\panel{f} that the impurity-substrate coupling reduces upon increasing the conductance, we move from the screened spin regime across the QPT to the free spin regime as the tip approaches the sample. This is consistent with the evolution of the Josephson current from an in-phase superposition (zero junction) to an out-of-phase superposition ($\pi$ junction) as the conductance increases.

At the QPT, a system typically becomes very sensitive to external parameters, such as temperature. Here, we note that the width of the QPT step in Fig.\ \ref{fig:theory}\panel{d} depends only on temperature, but experiences no broadening from voltage noise. This is in contrast to conventional scanning tunneling spectroscopy, where temperature broadening is typically obscured by voltage noise as well as interactions with the environment \cite{ast_sensing_2016}. Hence, YSR-tip functionalization may open new developments for low temperature thermomentry with high spatial resolution where measuring the slope of the QPT step accesses the temperature.

In summary, the experimental results directly reveal the consequences of the discrete parity change across the QPT in YSR states as well as the role of the impurity spin, which manifests itself in the supercurrent reversal. Our results establish an important connection to mesoscopic $\pi$-junctions providing the perspective to transfer some of their concepts, for example as sensing tools, to the atomic scale. Having direct tunable access to the QPT could be exploited to enhance the sensitivity in quantum sensing applications, such as a local temperature measurement. Also, demonstrating the coherent superposition of different transport channels in the DCB regime introduces a rudimentary phase sensitivity in STM measurements that can be exploited in other scenarios as well.

\section*{Acknowledgments}
We gratefully acknowledge stimulating discussions with Robert Drost, Berthold Jäck, and Francesco Tafuri. We dedicate this manuscript to the memory of Fabien Portier for the many inspirational discussions that the authors had with him. This work was funded in part by the ERC Consolidator Grant AbsoluteSpin (Grant No.\ 681164) and by the Center for Integrated Quantum Science and Technology (IQ$^\textrm{\small ST}$). J.A. acknowledges funding from the DFG under grant number AN336/11-1. C.P. acknowledges funding from the IQ$^\textrm{\small ST}$ and the Zeiss Foundation. A.L.Y. and J.C.C. acknowledge funding from the Spanish MINECO (Grant No. FIS2017-84057-P and FIS2017-84860-R), from the “Mar\'{\i}a de Maeztu” Programme for Units of Excellence in R\&D (MDM-2014-0377).

\clearpage
\newpage

\onecolumngrid
\begin{center}
\textbf{\large Supplementary Material for \\ Superconducting Quantum Interference at the Atomic Scale}
\end{center}
\vspace{1cm}
\twocolumngrid

\setcounter{figure}{0}
\setcounter{table}{0}
\setcounter{equation}{0}
\renewcommand{\thefigure}{S\arabic{figure}}
\renewcommand{\thetable}{S\Roman{table}}
\renewcommand{\theequation}{S\arabic{equation}}

\vspace{0.5cm}

\section{Materials and Methods}

Experiments were performed on Josephson nano-junctions built in a low temperature scanning tunneling microscope (STM) operated at 10\,mK. Approaching a superconducting vanadium tip, tailored with a \spinhalf\ impurity at its apex, to a crystalline V(100) substrate, we drove the impurity induced YSR states across the quantum phase transition (QPT) and detected a reversal of the supercurrent.

The V(100) substrate was cleaned by repeated cycles of Ar ion sputtering, annealing to $\sim925$\,K, and cooling to ambient temperature at a rate of 1-2\,K/s. Oxygen diffused from the bulk to the surface lead to typical surface reconstructions \cite{si_koller_structure_2001,si_kralj_hraes_2003}, which did not influence the characteristics of superconducting vanadium.
Surface defects mostly involve missing oxygen within the reconstruction, which appeared bright in STM topographs \cite{si_huang_tunnelling_2020}. Magnetic defects were found exhibiting YSR states at arbitrary energies within the gap as reported in Ref.\ \cite{si_huang_tunnelling_2020}.

The tip was sputtered in ultra-high vacuum and treated with field emission as well as subsequent indentation into the vanadium substrate until the expected gap of bulk vanadium appeared in the conductance spectrum. YSR tips were designed following the method of random dipping explained in Ref.\ \cite{si_huang_tunnelling_2020}. We purposefully chose to use YSR tips for our experiment as it gave better stability of the junction at higher conductance. Moreover, it offered better flexibility in designing and defining the junction over magnetic surface defects, which were mostly found to have a spatial extent of around 1\,nm.

\section{The Josephson Current}

Calculating the Josephson current is typically based on the energy-phase relation $E(\varphi)$ at zero applied bias voltage. In the DCB regime, the charging energy of the tunnel junction dominates, such that tunneling is charge dominated and sequential. The measurement is typically voltage-biased and the Cooper pair tunneling relies on the exchange of energy with the environment. The corresponding Cooper pair transfer can be calculated by means of a Fourier transform of the energy-phase relation
\begin{equation}
	E(\varphi) = \sum_{m = -\infty}^{+\infty} E_m e^{i m \varphi},
	\label{eq:ft}
\end{equation}
where $e^{i m \varphi}$ corresponds to the charge transfer operator and $m$ is the number of Cooper pairs being transferred. The Fourier components $E_m$ from Eq.\ \ref{eq:ft} are used to calculate the Josephson current
\begin{equation}
I(V)=\frac{2\pi}{\hbar} \sum_{m=1}^{+\infty}|E_{m}|^{2}(2me)\left[P_{m}(m2eV)-P_{m}(-m2eV)\right],
\label{eq:josiv}
\end{equation}
where $P_{m}(E)$ is the probability to exchange energy $E$ with the environment during the tunneling process, when $m$ Cooper pairs are tunneling. It is defined as a generalized $P(E)$-function \cite{si_senkpiel_single_2020}
\begin{equation}
P_{m}(E)=\int_{-\infty}^{+\infty}\frac{\mathrm{d}t}{2\pi\hbar}e^{m^{2}J(t)+iEt/\hbar}
\end{equation}
with the phase correlation function
\begin{equation}
J(t)=\braket{[\tilde{\varphi}(t)-\tilde{\varphi}(0)]\tilde{\varphi}(0)},
\end{equation}
accounting for Cooper-pair--phase fluctuations $\tilde{\varphi}=\varphi-2eVt/\hbar$ around the mean value determined by the external voltage. For a more detailed discussion see Ref.\ \cite{si_senkpiel_single_2020}.

\section{Derivation of the Energy-Phase Relation}

To describe the phase dependence of the energy of the YSR states we follow Ref.\ \cite{si_villas_interplay_2020,si_huang_quantum_2020} and make use of a mean-field Anderson model where a magnetic impurity coupled to superconducting leads is described by the following Hamiltonian
\begin{equation}
\label{eq-total-H}
H = H_{\rm t} + H_{\rm s} + H_{\rm i} + H_\mathrm{hopping} .
\end{equation}
Here, $H_j$, with $j={\rm t,s}$ (t stands for tip and s for substrate), is the BCS Hamiltonian of the lead $j$ given by
\begin{multline}
H_j = \sum_{{\boldsymbol k} \sigma} \xi_{{\boldsymbol k}j}
c^{\dagger}_{{\boldsymbol k}j\sigma} c_{{\boldsymbol k}j\sigma} \\
+ \sum_{\boldsymbol k} \left( \Delta_j e^{i\varphi_j}
c^{\dagger}_{{\boldsymbol k}j\uparrow} c^{\dagger}_{{-\boldsymbol k}j\downarrow}
+ \Delta_j e^{-i\varphi_j} c_{{-\boldsymbol k}j\downarrow} c_{{\boldsymbol k}j\uparrow}
\right) ,
\end{multline}
where $c^{\dagger}_{{\boldsymbol k}j\sigma}$ and $c_{{\boldsymbol k}j\sigma}$ are the creation and annihilation operators, respectively, of an electron of momentum ${\boldsymbol k}$, energy $\xi_{{\boldsymbol k}j}$, and spin $\sigma=\uparrow, \downarrow$ in lead $j$, $\Delta_j$ is the superconducting gap parameter, and $\varphi_j$ is the corresponding superconducting phase. On the other hand, $H_{\rm i}$ is the Hamiltonian of the magnetic impurity, which is given by
\begin{equation}
H_{\rm i} = E_\mathrm{U} (n_{\uparrow} + n_{\downarrow}) +
            E_\mathrm{J} (n_{\uparrow} - n_{\downarrow}) ,
\end{equation}
where $n_{\sigma} = d^{\dagger}_{\sigma} d_{\sigma}$ is the occupation number operator on the impurity, $E_{\rm U}$ is the on-site energy, and $E_{\rm J}$ is the exchange energy that breaks the spin degeneracy on the impurity. Finally, $H_\mathrm{hopping}$ describes the coupling between the magnetic impurity and the leads that adopts the form
\begin{equation}
H_{\rm hopping} = \sum_{{\boldsymbol k},j,\sigma} t_j \left(d^{\dagger}_{\sigma}
c_{{\boldsymbol k}j\sigma} + c^{\dagger}_{{\boldsymbol k}j\sigma} d_{\sigma}
\right) ,
\end{equation}
where $t_j$ describes the tunneling coupling between the impurity and the lead $j=\mathrm{t,s}$ and it is chosen to be real.

Now, it is convenient to rewrite the previous Hamiltonian in terms of four-dimensional spinors in a space resulting from the direct product of the spin space and the Nambu (electron-hole) space. In the case of the leads, the relevant spinor is defined as
\begin{equation}
\label{eq-c-tilde}
\tilde c^{\dagger}_{{\boldsymbol k}j} = \left( c^{\dagger}_{{\boldsymbol k}j\uparrow},
c_{{-\boldsymbol k}j\downarrow}, c^{\dagger}_{{\boldsymbol k}j\downarrow} ,
-c_{{-\boldsymbol k}j\uparrow} \right) ,
\end{equation}
while for the impurity states we define
\begin{equation}
\label{eq-d-tilde}
\tilde d^{\dagger} = \left( d^{\dagger}_{\uparrow}, d_{\downarrow},
d^{\dagger}_{\downarrow}, -d_{\uparrow} \right) .
\end{equation}

Using the notation $\tau_i$ and $\sigma_i$ ($i=1,2,3$) for Pauli matrices in Nambu and spin space, respectively, and with $\tau_0$ and $\sigma_0$ as the unit matrices in those spaces, it is easy to show that the Hamiltonian in Eq.~(\ref{eq-total-H}) can be cast into the form
\begin{subequations}
\begin{eqnarray}
H_j & = & \frac{1}{2} \sum_{\boldsymbol k} \tilde c^{\dagger}_{{\boldsymbol k}j}
\hat H_{{\boldsymbol k} j} \tilde c_{{\boldsymbol k}j} , \\
H_{\rm i} & = & \frac{1}{2} \tilde d^{\dagger} \hat H_{\rm i} \tilde d ,	 \\
H_{\rm hopping} & = & \frac{1}{2} \sum_{{\boldsymbol k},j} \left\{
\tilde c^{\dagger}_{{\boldsymbol k}j} \hat V_{j{\rm i}} \tilde d +
\tilde d^{\dagger} \hat V_{{\rm i}j} \tilde c_{{\boldsymbol k}j} \right\} ,
\end{eqnarray}
\end{subequations}
where
\begin{subequations}
\begin{eqnarray}
\hat H_{{\boldsymbol k} j} & = & \sigma_0 \otimes (\xi_{\boldsymbol k} \tau_3 +
\Delta_j e^{i \varphi_j \tau_3} \tau_1 ), \\
\hat H_{\rm i} & = & E_{\rm U} (\sigma_0 \otimes \tau_3) + E_{\rm J} (\sigma_3 \otimes \tau_0) , \\
\hat V_{j{\rm i}} & = & t_j(\sigma_0 \otimes \tau_3) =
\hat V^{\dagger}_{{\rm i}j} .
\end{eqnarray}
\end{subequations}

The starting point for the description of the electronic structure of our impurity system are the Green's functions of the different subsystems which can be easily calculated from the previous matrix Hamiltonians as follows. First, the retarded and advanced Green's functions of the leads resolved in $\boldsymbol k$-space are defined as $\hat g^\mathrm{r,a}_{{\boldsymbol k},jj}(E) = (E \pm i \eta - \hat H_{{\boldsymbol k} j})^{-1}$, where $j= {\rm t, s}$ and $\eta=0^+$ is a positive infinitesimal parameter, which we shall drop out to simplify things along with the superscript $\mathrm{r,a}$, unless they are strictly necessary. Summing over ${\boldsymbol k}$, $\hat g_{jj} = \sum_{\boldsymbol k} \hat g_{{\boldsymbol k},jj} \rightarrow \rho_{j}(0) \int^{\infty}_{-\infty} d \xi_{\boldsymbol k} \hat g_{jj}(\xi_{\boldsymbol k})$, where $\rho_{j}(0)$ is the normal density of states at the Fermi energy ($E=0$) of lead $j$, we arrive at the standard expression for the bulk Green's function of a BCS superconductor
\begin{equation}
\label{eq-gBCS}
\hat g_{jj}(E) = \frac{-\pi \rho_{j}(0)}{\sqrt{\Delta^2_j - E^2}} \sigma_0 \otimes\left[E \tau_0 +
\Delta_j e^{-i \varphi_j \tau_3} \tau_1 \right] .
\end{equation}

On the other hand, the impurity Green's function is given by
\begin{eqnarray}
\label{eq-gii}
\hat g_{\rm i i}(E) & = & (E- \hat H_{\rm i})^{-1}  \\
& = & \left( \begin{array}{cccc}
\frac{1}{E-E_{\rm U}-E_{\rm J}} & 0 & 0 & 0 \\ 0 & \frac{1}{E+E_{\rm U}-E_{\rm J}} & 0 & 0 \\
0 & 0 & \frac{1}{E-E_{\rm U}+E_{\rm J}} & 0 \\ 0 & 0 & 0 & \frac{1}{E+E_{\rm U}+E_{\rm J}}
\nonumber \end{array} \right).	
\end{eqnarray}

The dressed Green's functions of the impurity system taking into account the coupling to the superconducting leads are calculated by solving the Dyson equations
\begin{equation}
\hat G_{jk} = \hat g_{jk} + \sum_{\alpha \beta} \hat g_{j \alpha}
\hat V_{\alpha \beta} \hat G_{\beta k} ,	
\end{equation}
where the indices run over ${\rm t}$, ${\rm i}$ and ${\rm s}$ and the couplings are given by $\hat V_{j{\rm i}} = t_{j} (\sigma_0 \otimes \tau_3) = \hat V^{\dagger}_{{\rm i}j}$. Assuming equal gaps in both superconducting leads ($\Delta_{\rm t} = \Delta_{\rm s} =\Delta=725\,\upmu$eV), these equations can be easily solved to obtain the following dressed Green's functions on the impurity site
\begin{equation}
\label{eq-Gii}
\hat G_{\rm ii}(E) = \left( \begin{array}{cc} \hat G_{{\rm ii},\uparrow \uparrow}(E) & 0 \\
0 & \hat G_{{\rm ii},\downarrow \downarrow}(E) \end{array} \right) ,
\end{equation}
with
\begin{widetext}
\begin{equation}
\hat G_{{\rm ii},\sigma \sigma}(E) = \frac{1}{D_{\sigma}(E)} \left( \begin{array}{cc}
E (\Gamma_{\rm s} + \Gamma_{\rm t}) + (E+E_{\rm U}-E_{\rm J \sigma}) \sqrt{\Delta^2 - E^2} &
\Gamma_{\rm s} \Delta e^{i \varphi_{\rm s}} +
\Gamma_{\rm t} \Delta e^{i \varphi_{\rm t}} \\
\Gamma_{\rm s} \Delta e^{- i \varphi_{\rm s}} +
\Gamma_{\rm t} \Delta e^{-i \varphi_{\rm t}} &
E (\Gamma_{\rm s} + \Gamma_{\rm t})  + (E-E_{\rm U}-E_{\rm J \sigma}) \sqrt{\Delta^2 - E^2}
\end{array} \right) ,
\end{equation}
where
\begin{equation}
\label{eq-D}
D_{\sigma}(E,\varphi) = 2(\Gamma_{\rm s} + \Gamma_{\rm t}) E (E-E_{\rm J \sigma}) +
\left[(E-E_{\rm J \sigma})^2 - E^2_{\rm U} - (\Gamma_{\rm s} + \Gamma_{\rm t})^2 \right] \sqrt{\Delta^2 - E^2} +
\frac{4\Delta^2}{\sqrt{\Delta^2 - E^2}} \Gamma_{\rm s} \Gamma_{\rm t} \sin^2(\varphi/2) .
\end{equation}
\end{widetext}
Here, $\varphi = \varphi_{\rm t} - \varphi_{\rm S}$ is the superconducting phase difference, $E_{\rm J \uparrow} = + E_{\rm J}$ and $E_{\rm J \downarrow} = -E_{\rm J}$, and we have defined the tunneling rates $\Gamma_{j} = \pi \rho_{j}(0) t^2_{j}$ with $j={\rm t,s}$.

As usual, the local density of states (LDOS) projected onto the impurity site can be obtained from the imaginary part of diagonal elements of the Green's function above. The result for the total LDOS adopts the following form
\begin{equation}
\label{eq-LDOS-imp}	
\rho_{\rm Total,imp}(E,\varphi) = \rho_{\uparrow}(E,\varphi) + \rho_{\downarrow}(E,\varphi),
\end{equation}
where
\begin{multline}
\label{eq-LDOS-imp-sigma}
\rho_{\sigma}(E,\varphi) = \\ \frac{1}{\pi} \mbox{Im} \left\{ \frac{E (\Gamma_{\rm s} + \Gamma_{\rm t}) +
(E+E_{\rm U}-E_{\rm J \sigma}) \sqrt{\Delta^2 - E^2}}{D_{\sigma}(E,\varphi)} \right\} , 	
\end{multline}
where $D_{\sigma}(E,\varphi)$ is given by Eq.~(\ref{eq-D}). The condition for the appearance of superconducting bound states is $D_{\sigma}(E,\varphi) = 0$. In particular, the spin-induced YSR states appear in the limit $E_{\rm J} \gg \Delta$ (and they are inside the gap when also $\Gamma_\mathrm{t} \gg \Delta$). Additionally, we focus on the tunnel regime where $\Gamma_\mathrm{s} \ll \Gamma_\mathrm{t}$, which is the relevant regime for our experiments. Note that in our experiment the impurity is strongly coupled to the tip (t) and tunneling is between the impurity and the substrate (s). The phase-dependent energies of the YSR bound states are approximately given by
\begin{multline}
\label{eq-YSR}
E^{\rm YSR}(\varphi) = \\ \pm \Delta \frac{ (\Gamma_{\rm s} + \Gamma_{\rm t})^2 + E_{\rm U}^2 -
E_{\rm J}^2 - 4 \Gamma_{\rm s} \Gamma_{\rm t} \sin^2(\varphi/2)}
{\sqrt{ \left[(E_{\rm U}-E_{\rm J})^2 +(\Gamma_{\rm s} + \Gamma_{\rm t})^2 \right]
\left[(E_{\rm U}+E_{\rm J})^2 +(\Gamma_{\rm s} + \Gamma_{\rm t})^2 \right]}} .
\end{multline}
The transmission $\tau_\text{YSR}$ through the YSR channel is given by
\begin{equation}
    \tau_\text{YSR} = \frac{4\Gamma_\text{s}\Gamma_\text{t}}{(E_\text{U} - E_\text{J})^2 + (\Gamma_\text{s} + \Gamma_\text{t})^2} + \frac{4\Gamma_\text{s}\Gamma_\text{t}}{(E_\text{U} + E_\text{J})^2 + (\Gamma_\text{s} + \Gamma_\text{t})^2}.
\end{equation}

If the impurity is only coupled to one of the superconductors, the energy of the YSR state is
\begin{equation}
    \varepsilon = \pm \Delta \frac{ \Gamma_{j}^2 + E_{\rm U}^2 -
E_{\rm J}^2}
{\sqrt{ \left[(E_{\rm U}-E_{\rm J})^2 +\Gamma_{j}^2 \right]
\left[(E_{\rm U}+E_{\rm J})^2 + \Gamma_{j}^2 \right]}}.
\end{equation}
At the QPT, the energy of the YSR state crosses zero ($\varepsilon = 0$), which marks a circle in the phase diagram in Fig.\ 1\panel{b} of the main text according to
\begin{equation}
    E_{\rm J}^2 = \Gamma_{j}^2 + E_{\rm U}^2 = \text{const}.
\end{equation}

\begin{figure}
    \centering
    \includegraphics[width=\columnwidth]{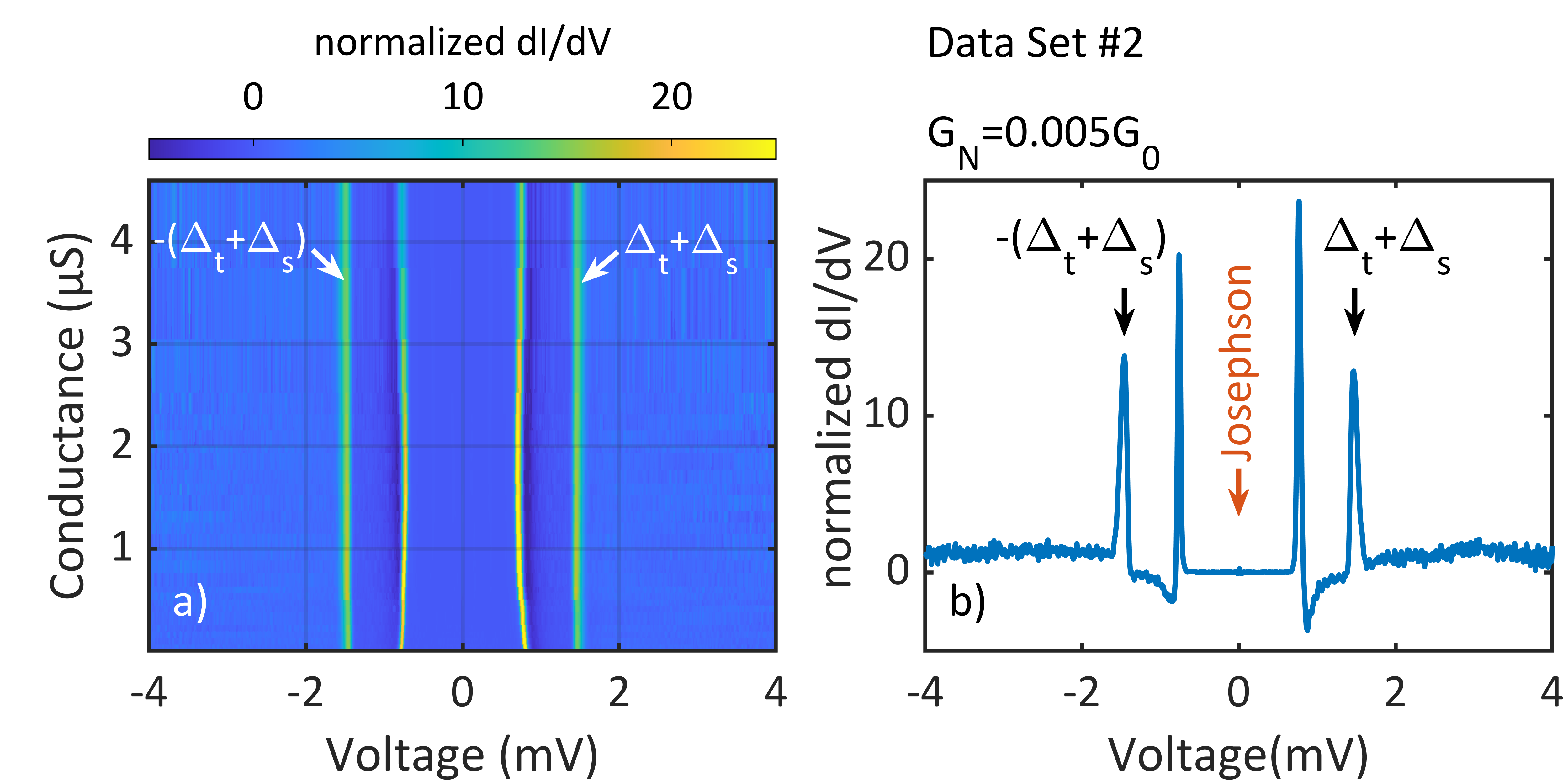}
    \caption{\textbf{Atomic YSR State (Data Set \#2)} \panelcaption{a} Differential conductance spectra as function of bias voltage ($x$-axis) and conductance ($y$-axis). The YSR states and the coherence peaks are nearly equally prominent. \panelcaption{b} Horizontal line cut through (d) to show the YSR peaks in relation to the coherence peaks, which are more prominent in this data set compared to the data in the main text. This indicates that the transmission through the BCS reference channel is higher than for the data in the main text.}
    \label{fig:dataset2}
\end{figure}

\begin{SCfigure*}
    \centering
    \includegraphics[width=0.6\textwidth]{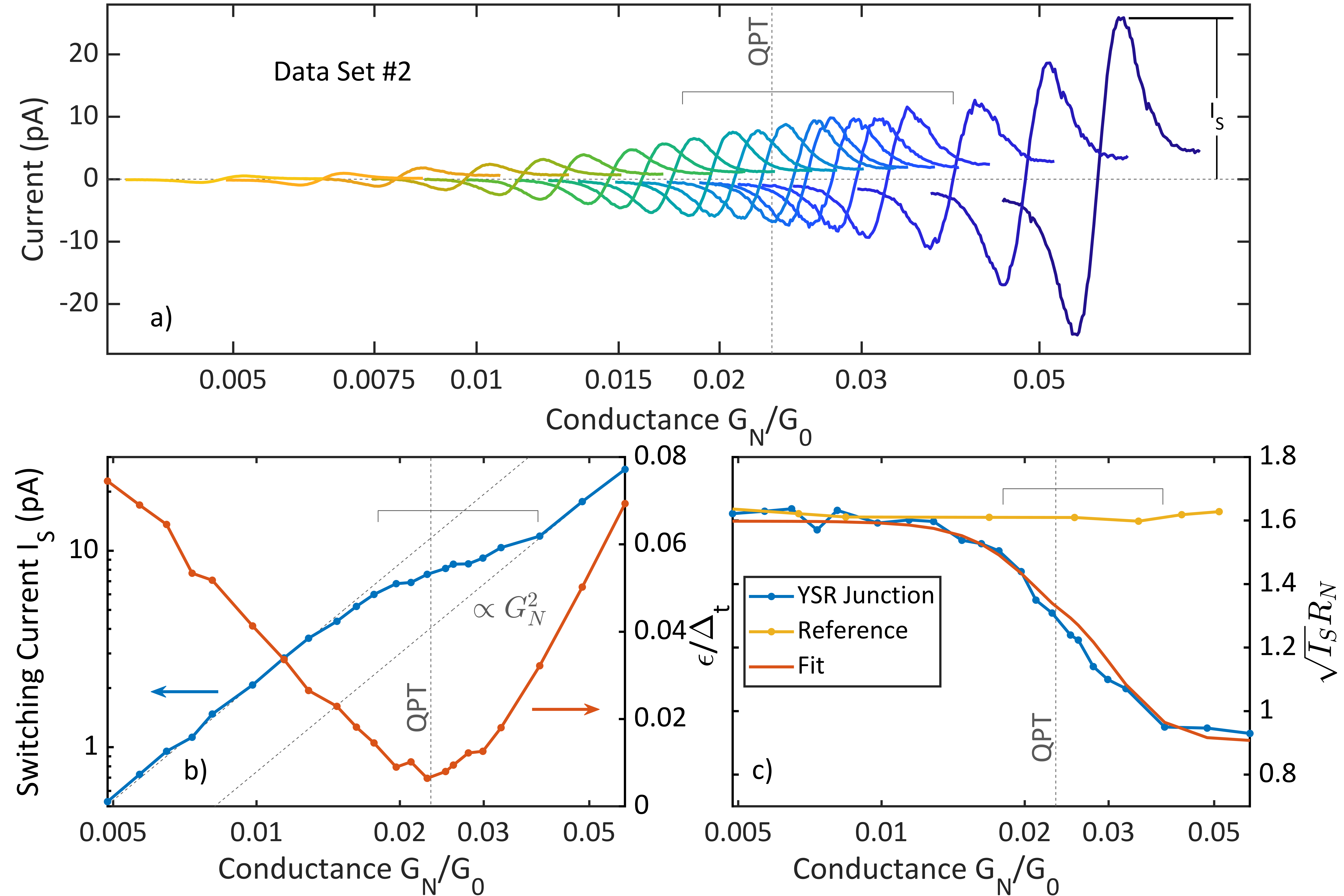}
    \caption{\textbf{Josephson Effect (Data Set \#2)} \panelcaption{a} Josephson spectra $I(V)$ in a range of $\pm 60\,\upmu$eV shifted horizontally by the conductance at which they were measured. The horizontal bracket indicates the region, where the evolution deviates from the conventional Ambegaokar-Baratoff formula. The quantum phase transition (QPT) is indicated by the vertical dashed line. \panelcaption{b} The switching current $I_\text{S}$, which is the current maximum indicated in \panelsubcaption{a} as function of the normal state conductance $G_\text{N}$ (blue line). The square dependence at low and high conductance is indicated by dashed lines (labeled by $\propto G_\text{N}^2$). The YSR energy as function of conductance is shown as a blue line. The minimum indicates the QPT (vertical dashed line). \panelcaption{c} This graph shows $\sqrt{I_\text{S}}R_\text{N}$ (blue line) of the data in \panelsubcaption{b}, which is proportional to $I_\text{C}R_\text{N}$ for a harmonic energy-phase relation ($R_\text{N}$: renormalized resistance (see text), $I_\text{C}$: critical current). A reference junction without any YSR states is shown as a yellow line indicating the expected evolution according to the Ambegaokar-Baratoff formula.}
    \label{fig:analysisset2}
\end{SCfigure*}

\section{The Energy-Phase Relation}

The energy-phase relation for a tunnel junction with two equal conventional Bardeen-Cooper-Schrieffer (BCS) gaps (i.e.\ $\Delta_\text{t} = \Delta_\text{s} = \Delta$) in tip ($t$) and sample ($s$) is
\begin{equation}
	E^\text{BCS}(\varphi) = \pm \Delta \sqrt{1 - \tau_\text{BCS} \sin^2 \left( \frac{\varphi}{2} \right) },
	\label{eq:absfull}
\end{equation}
where $\tau_\text{BCS}$ is the BCS channel transmission. For small channel transmission ($\tau_\text{BCS}\ll 1$), the energy-phase relation becomes harmonic and the amplitude of the first harmonic ($m=1$) following Eq.\ \eqref{eq:ft} is
\begin{equation}
	E^\text{BCS}_\text{1} = \frac{\Delta}{4}\tau_\text{BCS},
	\label{eq:ejbcs}
\end{equation}

The energy-phase relation of a tunnel junction with one YSR state is
given in Eq.\ \eqref{eq-YSR}. For the YSR junction, we expect higher harmonic contributions only very close to the QPT. We can explain all experimental data based on the first harmonic contribution, so we will neglect higher order contributions here. The amplitude in the lowest harmonic ($m=1$) for Eq.\ \eqref{eq-YSR} is
\begin{equation}
    E^\text{YSR}_\text{1} = \frac{\text{sgn}[(\Gamma_{\rm s} + \Gamma_{\rm t})^2 + E_{\rm U}^2 -
E_{\rm J}^2]\ 2 \Delta \Gamma_\text{s}\Gamma_\text{t}}{\sqrt{\left[(E_\text{U} - E_\text{J})^2 + (\Gamma_\text{s} + \Gamma_\text{t})^2\right] \left[(E_\text{U} + E_\text{J})^2 + (\Gamma_\text{s} + \Gamma_\text{t})^2\right]}},
\end{equation}
where sgn is the sign function. The only way that we can model the experimentally observed step is by coherently adding the two transport channels
\begin{equation}
    E(\varphi) = E^\text{BCS}(\varphi) + E^\text{YSR}(\varphi)
\end{equation}
and by extension the two harmonic amplitudes, which enter into the calculation of the Josephson current in Eq.\ \ref{eq:josiv}
\begin{equation}
    E_1 = E^\text{BCS}_1 + E^\text{YSR}_1.
    \label{eq:sumharm}
\end{equation}
Equation \eqref{eq:sumharm} is valid at zero temperature. At finite temperature, thermal excitations have to be taken into account.

\section{Thermal Excitation of the YSR State}

In the vicinity of the QPT, the energy of the YSR state become so small that it may be thermally excited. This probability has to be taken into account when calculating the Josephson current as it can significantly broaden the region of the QPT even at mK temperatures. As the energy of the Andreev bound state in the BCS channel is $E^\text{BCS}>\Delta\sqrt{1-\tau}$, it is unlikely for this state to be thermally excited. We, therefore, only consider the thermal excitation of the YSR state. Here, the energy of the excited state is $-E^{\text{YSR}}(\varphi)$. If we now define $p$ as the probability for the YSR state to be thermally excited, we can write the lowest order harmonic coefficient as
\begin{equation}
    E_1^2 = (1-p)\left|E^\text{BCS}_1+E^\text{YSR}_1\right|^2 + p\left|E^\text{BCS}_1-E^\text{YSR}_1\right|^2,
    \label{eq:coeff}
\end{equation}
where $p=2/(1+\exp(|\varepsilon|/k_\text{B}T))$, $\varepsilon$ is the energy of the YSR state, and $T$ is the temperature. We use the coefficient in Eq.\ \ref{eq:coeff} to calculate the Josephson current in Eq.\ \ref{eq:josiv}, which is used to calculate the fit in the main text.

\section{The $I_\text{C}R_\text{N}$ product for a YSR state}

In the presence of a YSR state, the normal state resistance $R_\text{N}$, which is relevant for modeling the Josephson current may be modified in the presence of Kondo correlations resulting from an enhanced density of states in the range of the superconducting gap. Therefore, the normal state conductance $G_\text{N}$ that is typically measured outside the superconducting gap has to be renormalized as the assumption that the density of states is a constant no longer holds. We have used reference measurements through a tunnel junction without any YSR state \cite{si_jack_critical_2016}, which has been shown previously to follow the expected Ambegaokar-Baratoff formula \cite{si_ambegaokar_tunneling_1963}, to find the renormalization coefficient for the normal state resistance $R_\text{N}$ in the presence of a YSR state. For the data set presented in the main text, we find $R_\text{N}=1/(2.05 G_\text{N})$, where $G_\text{N}$ is the normal state conductance outside the superconducting gap. We note that this is a phenomenological result. The value of the renormalization factor only holds for this particular data set as the contributions from Kondo correlations to the local density of states may vary between impurities.

\section{Other YSR Tips}

In Fig.\ \ref{fig:dataset2}, we show another data set with an STM tip that is functionalized with a YSR state. As in the main text, the YSR state energy changes as a function of conductance (i.e.\ tip-sample distance) as shown in Fig.\ \ref{fig:dataset2}\panel{a}. The YSR state undergoes a QPT at $0.023G_0$ from a screened spin to a free spin regime (for comparison, the data in the main text has the QPT at $0.009G_0$). A spectrum near the QPT is shown in Fig.\ \ref{fig:dataset2}\panel{b} to show the transmission ratio between the transport channel through the YSR state and the reference BCS gap. We can clearly see sizeable coherence peaks at $\pm(\Delta_\text{t}+\Delta_\text{s})$ indicating that the reference BCS channel is much stronger than the YSR channel ($\Delta_\text{t}=\Delta_\text{s}=715\,\upmu$eV).

The Josephson effect can be seen throughout the range of conductance values shown in Fig.\ \ref{fig:dataset2}. Figure \ref{fig:analysisset2} shows the analysis of the Josephson effect for the second data set in analogy to Fig.\ 2 of the main text. In Fig.\ \ref{fig:analysisset2}\panel{a}, we see the evolution of the Josephson spectra as function of conductance. Each spectrum is shown in a range of $\pm60\,\upmu$eV and horizontally offset by the corresponding conductance value. We can clearly see an increase in the signal, but which is again modulated in the region of the square bracket in Fig.\ \ref{fig:analysisset2}\panel{a}. We plot the switching current $I_\text{S}$ (as indicated in Fig.\ \ref{fig:analysisset2}\panel{a}) as a function of conductance in Fig.\ \ref{fig:analysisset2}\panel{b} (blue line). The corresponding YSR energies are plotted as a red line. We indicate the position of the QPT by a vertical dashed line, where the YSR energy has a minimum. The switching current also changes in the region of the QPT as in Fig.\ 2 of the main text.

In Fig.\ \ref{fig:analysisset2}\panel{c}, the $\sqrt{I_\text{S}}R_\text{N}$ product of the experimental data is plotted (blue line) in relation to a reference junction (yellow line) and a fit (orange line). For this data set, the renormalization of the normal state resistance $R^\text{YSR}_\text{N}=1/(1.19 G_\text{N})$ through the YSR state is not as pronoounced because the transmission through the YSR channel is much reduced compared to the data set in the main text. For the fit, we use the same effective temperature of $T_\text{eff}=75\,$mK as for the data in the main text. For the relative channel transmission, we find 77.7\% and 22.3\% for the BCS and the YSR channel, respectively. The overall agreement between experimental data and fit is excellent.


\begin{thebibliography}{43}%
\makeatletter
\providecommand \@ifxundefined [1]{%
 \@ifx{#1\undefined}
}%
\providecommand \@ifnum [1]{%
 \ifnum #1\expandafter \@firstoftwo
 \else \expandafter \@secondoftwo
 \fi
}%
\providecommand \@ifx [1]{%
 \ifx #1\expandafter \@firstoftwo
 \else \expandafter \@secondoftwo
 \fi
}%
\providecommand \natexlab [1]{#1}%
\providecommand \emph  [1]{``#1''}%
\providecommand \bibnamefont  [1]{#1}%
\providecommand \bibfnamefont [1]{#1}%
\providecommand \citenamefont [1]{#1}%
\providecommand \href@noop [0]{\@secondoftwo}%
\providecommand \href [0]{\begingroup \@sanitize@url \@href}%
\providecommand \@href[1]{\@@startlink{#1}\@@href}%
\providecommand \@@href[1]{\endgroup#1\@@endlink}%
\providecommand \@sanitize@url [0]{\catcode `\\12\catcode `\$12\catcode
  `\&12\catcode `\#12\catcode `\^12\catcode `\_12\catcode `\%12\relax}%
\providecommand \@@startlink[1]{}%
\providecommand \@@endlink[0]{}%
\providecommand \url  [0]{\begingroup\@sanitize@url \@url }%
\providecommand \@url [1]{\endgroup\@href {#1}{\urlprefix }}%
\providecommand \urlprefix  [0]{URL }%
\providecommand \Eprint [0]{\href }%
\providecommand \doibase [0]{http://dx.doi.org/}%
\providecommand \selectlanguage [0]{\@gobble}%
\providecommand \bibinfo  [0]{\@secondoftwo}%
\providecommand \bibfield  [0]{\@secondoftwo}%
\providecommand \translation [1]{[#1]}%
\providecommand \BibitemOpen [0]{}%
\providecommand \bibitemStop [0]{}%
\providecommand \bibitemNoStop [0]{.\EOS\space}%
\providecommand \EOS [0]{\spacefactor3000\relax}%
\providecommand \BibitemShut  [1]{\csname bibitem#1\endcsname}%
\let\auto@bib@innerbib\@empty
\bibitem [{\citenamefont {Josephson}(1962)}]{josephson_possible_1962}%
  \BibitemOpen
  \bibfield  {author} {\bibinfo {author} {\bibfnamefont {B.~D.}\ \bibnamefont
  {Josephson}},\ }\bibfield  {title} {\emph {\bibinfo {title} {Possible new
  effects in superconductive tunnelling},}\ }\href {\doibase
  10.1016/0031-9163(62)91369-0} {\bibfield  {journal} {\bibinfo  {journal}
  {Physics Letters}\ }\textbf {\bibinfo {volume} {1}},\ \bibinfo {pages} {251}
  (\bibinfo {year} {1962})}\BibitemShut {NoStop}%
\bibitem [{\citenamefont {Kulik}(1966)}]{kulik_magnitude_1966}%
  \BibitemOpen
  \bibfield  {author} {\bibinfo {author} {\bibfnamefont {I.}~\bibnamefont
  {Kulik}},\ }\bibfield  {title} {\emph {\bibinfo {title} {Magnitude of the
  {Critical} {Josephson} {Tunnel} {Current}},}\ }\href
  {http://www.jetp.ac.ru/cgi-bin/e/index/e/22/4/p841?a=list} {\bibfield
  {journal} {\bibinfo  {journal} {Journal of Experimental and Theoretical
  Physics}\ }\textbf {\bibinfo {volume} {22}},\ \bibinfo {pages} {841}
  (\bibinfo {year} {1966})}\BibitemShut {NoStop}%
\bibitem [{\citenamefont {Tsuei}\ \emph {et~al.}(1994)\citenamefont {Tsuei},
  \citenamefont {Kirtley}, \citenamefont {Chi}, \citenamefont {Yu-Jahnes},
  \citenamefont {Gupta}, \citenamefont {Shaw}, \citenamefont {Sun},\ and\
  \citenamefont {Ketchen}}]{tsuei_pairing_1994}%
  \BibitemOpen
  \bibfield  {author} {\bibinfo {author} {\bibfnamefont {C.~C.}\ \bibnamefont
  {Tsuei}}, \bibinfo {author} {\bibfnamefont {J.~R.}\ \bibnamefont {Kirtley}},
  \bibinfo {author} {\bibfnamefont {C.~C.}\ \bibnamefont {Chi}}, \bibinfo
  {author} {\bibfnamefont {L.~S.}\ \bibnamefont {Yu-Jahnes}}, \bibinfo {author}
  {\bibfnamefont {A.}~\bibnamefont {Gupta}}, \bibinfo {author} {\bibfnamefont
  {T.}~\bibnamefont {Shaw}}, \bibinfo {author} {\bibfnamefont {J.~Z.}\
  \bibnamefont {Sun}}, \ and\ \bibinfo {author} {\bibfnamefont {M.~B.}\
  \bibnamefont {Ketchen}},\ }\bibfield  {title} {\emph {\bibinfo {title}
  {Pairing {Symmetry} and {Flux} {Quantization} in a {Tricrystal}
  {Superconducting} {Ring} of {YBa$_2$Cu$_3$O$_{7-\delta}$}},}\ }\href
  {\doibase 10.1103/PhysRevLett.73.593} {\bibfield  {journal} {\bibinfo
  {journal} {Physical Review Letters}\ }\textbf {\bibinfo {volume} {73}},\
  \bibinfo {pages} {593} (\bibinfo {year} {1994})}\BibitemShut {NoStop}%
\bibitem [{\citenamefont {Kirtley}\ \emph {et~al.}(1995)\citenamefont
  {Kirtley}, \citenamefont {Tsuei}, \citenamefont {Sun}, \citenamefont {Chi},
  \citenamefont {Yu-Jahnes}, \citenamefont {Gupta}, \citenamefont {Rupp},\ and\
  \citenamefont {Ketchen}}]{kirtley_symmetry_1995}%
  \BibitemOpen
  \bibfield  {author} {\bibinfo {author} {\bibfnamefont {J.~R.}\ \bibnamefont
  {Kirtley}}, \bibinfo {author} {\bibfnamefont {C.~C.}\ \bibnamefont {Tsuei}},
  \bibinfo {author} {\bibfnamefont {J.~Z.}\ \bibnamefont {Sun}}, \bibinfo
  {author} {\bibfnamefont {C.~C.}\ \bibnamefont {Chi}}, \bibinfo {author}
  {\bibfnamefont {L.~S.}\ \bibnamefont {Yu-Jahnes}}, \bibinfo {author}
  {\bibfnamefont {A.}~\bibnamefont {Gupta}}, \bibinfo {author} {\bibfnamefont
  {M.}~\bibnamefont {Rupp}}, \ and\ \bibinfo {author} {\bibfnamefont {M.~B.}\
  \bibnamefont {Ketchen}},\ }\bibfield  {title} {{\selectlanguage
  {english}\emph {\bibinfo {title} {Symmetry of the order parameter in the
  high-{T$_\text{c}$} superconductor {YBa$_2$Cu$_3$O$_{7-\delta}$}},}\ }}\href
  {\doibase 10.1038/373225a0} {\bibfield  {journal} {\bibinfo  {journal}
  {Nature}\ }\textbf {\bibinfo {volume} {373}},\ \bibinfo {pages} {225}
  (\bibinfo {year} {1995})}\BibitemShut {NoStop}%
\bibitem [{\citenamefont
  {Van~Harlingen}(1995)}]{harlingen_phase-sensitive_1995}%
  \BibitemOpen
  \bibfield  {author} {\bibinfo {author} {\bibfnamefont {D.~J.}\ \bibnamefont
  {Van~Harlingen}},\ }\bibfield  {title} {\emph {\bibinfo {title}
  {Phase-sensitive tests of the symmetry of the pairing state in the
  high-temperature superconductors---{Evidence} for $d_{x^2-y^2}$ symmetry},}\
  }\href {\doibase 10.1103/RevModPhys.67.515} {\bibfield  {journal} {\bibinfo
  {journal} {Reviews of Modern Physics}\ }\textbf {\bibinfo {volume} {67}},\
  \bibinfo {pages} {515} (\bibinfo {year} {1995})}\BibitemShut {NoStop}%
\bibitem [{\citenamefont {Wollman}\ \emph {et~al.}(1995)\citenamefont
  {Wollman}, \citenamefont {Van~Harlingen}, \citenamefont {Giapintzakis},\ and\
  \citenamefont {Ginsberg}}]{wollman_evidence_1995}%
  \BibitemOpen
  \bibfield  {author} {\bibinfo {author} {\bibfnamefont {D.~A.}\ \bibnamefont
  {Wollman}}, \bibinfo {author} {\bibfnamefont {D.~J.}\ \bibnamefont
  {Van~Harlingen}}, \bibinfo {author} {\bibfnamefont {J.}~\bibnamefont
  {Giapintzakis}}, \ and\ \bibinfo {author} {\bibfnamefont {D.~M.}\
  \bibnamefont {Ginsberg}},\ }\bibfield  {title} {\emph {\bibinfo {title}
  {Evidence for $d_{x^2-y^2}$ {Pairing} from the {Magnetic} {Field}
  {Modulation} of {YBa$_2$Cu$_3$O$_7$-Pb} {Josephson} {Junctions}},}\ }\href
  {\doibase 10.1103/PhysRevLett.74.797} {\bibfield  {journal} {\bibinfo
  {journal} {Physical Review Letters}\ }\textbf {\bibinfo {volume} {74}},\
  \bibinfo {pages} {797} (\bibinfo {year} {1995})}\BibitemShut {NoStop}%
\bibitem [{\citenamefont {Tsuei}\ and\ \citenamefont
  {Kirtley}(2000)}]{tsuei_pairing_2000}%
  \BibitemOpen
  \bibfield  {author} {\bibinfo {author} {\bibfnamefont {C.~C.}\ \bibnamefont
  {Tsuei}}\ and\ \bibinfo {author} {\bibfnamefont {J.~R.}\ \bibnamefont
  {Kirtley}},\ }\bibfield  {title} {\emph {\bibinfo {title} {Pairing symmetry
  in cuprate superconductors},}\ }\href {\doibase 10.1103/RevModPhys.72.969}
  {\bibfield  {journal} {\bibinfo  {journal} {Rev. Mod. Phys.}\ }\textbf
  {\bibinfo {volume} {72}},\ \bibinfo {pages} {969} (\bibinfo {year}
  {2000})}\BibitemShut {NoStop}%
\bibitem [{\citenamefont {Gingrich}\ \emph {et~al.}(2016)\citenamefont
  {Gingrich}, \citenamefont {Niedzielski}, \citenamefont {Glick}, \citenamefont
  {Wang}, \citenamefont {Miller}, \citenamefont {Loloee}, \citenamefont
  {Pratt~Jr},\ and\ \citenamefont {Birge}}]{gingrich_controllable_2016}%
  \BibitemOpen
  \bibfield  {author} {\bibinfo {author} {\bibfnamefont {E.~C.}\ \bibnamefont
  {Gingrich}}, \bibinfo {author} {\bibfnamefont {B.~M.}\ \bibnamefont
  {Niedzielski}}, \bibinfo {author} {\bibfnamefont {J.~A.}\ \bibnamefont
  {Glick}}, \bibinfo {author} {\bibfnamefont {Y.}~\bibnamefont {Wang}},
  \bibinfo {author} {\bibfnamefont {D.~L.}\ \bibnamefont {Miller}}, \bibinfo
  {author} {\bibfnamefont {R.}~\bibnamefont {Loloee}}, \bibinfo {author}
  {\bibfnamefont {W.~P.}\ \bibnamefont {Pratt~Jr}}, \ and\ \bibinfo {author}
  {\bibfnamefont {N.~O.}\ \bibnamefont {Birge}},\ }\bibfield  {title}
  {{\selectlanguage {english}\emph {\bibinfo {title} {Controllable 0-$\pi$
  {Josephson} junctions containing a ferromagnetic spin valve},}\ }}\href
  {\doibase 10.1038/nphys3681} {\bibfield  {journal} {\bibinfo  {journal} {Nat.
  Phys.}\ }\textbf {\bibinfo {volume} {12}},\ \bibinfo {pages} {564} (\bibinfo
  {year} {2016})}\BibitemShut {NoStop}%
\bibitem [{\citenamefont {Cleuziou}\ \emph {et~al.}(2006)\citenamefont
  {Cleuziou}, \citenamefont {Wernsdorfer}, \citenamefont {Bouchiat},
  \citenamefont {Ondarçuhu},\ and\ \citenamefont
  {Monthioux}}]{cleuziou_carbon_2006}%
  \BibitemOpen
  \bibfield  {author} {\bibinfo {author} {\bibfnamefont {J.-P.}\ \bibnamefont
  {Cleuziou}}, \bibinfo {author} {\bibfnamefont {W.}~\bibnamefont
  {Wernsdorfer}}, \bibinfo {author} {\bibfnamefont {V.}~\bibnamefont
  {Bouchiat}}, \bibinfo {author} {\bibfnamefont {T.}~\bibnamefont
  {Ondarçuhu}}, \ and\ \bibinfo {author} {\bibfnamefont {M.}~\bibnamefont
  {Monthioux}},\ }\bibfield  {title} {{\selectlanguage {english}\emph {\bibinfo
  {title} {Carbon nanotube superconducting quantum interference device},}\
  }}\href {\doibase 10.1038/nnano.2006.54} {\bibfield  {journal} {\bibinfo
  {journal} {Nature Nanotechnology}\ }\textbf {\bibinfo {volume} {1}},\
  \bibinfo {pages} {53} (\bibinfo {year} {2006})}\BibitemShut {NoStop}%
\bibitem [{\citenamefont {Feofanov}\ \emph {et~al.}(2010)\citenamefont
  {Feofanov}, \citenamefont {Oboznov}, \citenamefont {Bol’ginov},
  \citenamefont {Lisenfeld}, \citenamefont {Poletto}, \citenamefont {Ryazanov},
  \citenamefont {Rossolenko}, \citenamefont {Khabipov}, \citenamefont
  {Balashov}, \citenamefont {Zorin}, \citenamefont {Dmitriev}, \citenamefont
  {Koshelets},\ and\ \citenamefont {Ustinov}}]{feofanov_implementation_2010}%
  \BibitemOpen
  \bibfield  {author} {\bibinfo {author} {\bibfnamefont {A.~K.}\ \bibnamefont
  {Feofanov}}, \bibinfo {author} {\bibfnamefont {V.~A.}\ \bibnamefont
  {Oboznov}}, \bibinfo {author} {\bibfnamefont {V.~V.}\ \bibnamefont
  {Bol’ginov}}, \bibinfo {author} {\bibfnamefont {J.}~\bibnamefont
  {Lisenfeld}}, \bibinfo {author} {\bibfnamefont {S.}~\bibnamefont {Poletto}},
  \bibinfo {author} {\bibfnamefont {V.~V.}\ \bibnamefont {Ryazanov}}, \bibinfo
  {author} {\bibfnamefont {A.~N.}\ \bibnamefont {Rossolenko}}, \bibinfo
  {author} {\bibfnamefont {M.}~\bibnamefont {Khabipov}}, \bibinfo {author}
  {\bibfnamefont {D.}~\bibnamefont {Balashov}}, \bibinfo {author}
  {\bibfnamefont {A.~B.}\ \bibnamefont {Zorin}}, \bibinfo {author}
  {\bibfnamefont {P.~N.}\ \bibnamefont {Dmitriev}}, \bibinfo {author}
  {\bibfnamefont {V.~P.}\ \bibnamefont {Koshelets}}, \ and\ \bibinfo {author}
  {\bibfnamefont {A.~V.}\ \bibnamefont {Ustinov}},\ }\bibfield  {title}
  {{\selectlanguage {english}\emph {\bibinfo {title} {Implementation of
  superconductor/ferromagnet/ superconductor $\pi$-shifters in superconducting
  digital and quantum circuits},}\ }}\href {\doibase 10.1038/nphys1700}
  {\bibfield  {journal} {\bibinfo  {journal} {Nature Physics}\ }\textbf
  {\bibinfo {volume} {6}},\ \bibinfo {pages} {593} (\bibinfo {year}
  {2010})}\BibitemShut {NoStop}%
\bibitem [{\citenamefont {Ryazanov}\ \emph {et~al.}(2001)\citenamefont
  {Ryazanov}, \citenamefont {Oboznov}, \citenamefont {Rusanov}, \citenamefont
  {Veretennikov}, \citenamefont {Golubov},\ and\ \citenamefont
  {Aarts}}]{ryazanov_coupling_2001}%
  \BibitemOpen
  \bibfield  {author} {\bibinfo {author} {\bibfnamefont {V.~V.}\ \bibnamefont
  {Ryazanov}}, \bibinfo {author} {\bibfnamefont {V.~A.}\ \bibnamefont
  {Oboznov}}, \bibinfo {author} {\bibfnamefont {A.~Y.}\ \bibnamefont
  {Rusanov}}, \bibinfo {author} {\bibfnamefont {A.~V.}\ \bibnamefont
  {Veretennikov}}, \bibinfo {author} {\bibfnamefont {A.~A.}\ \bibnamefont
  {Golubov}}, \ and\ \bibinfo {author} {\bibfnamefont {J.}~\bibnamefont
  {Aarts}},\ }\bibfield  {title} {\emph {\bibinfo {title} {Coupling of two
  superconductors through a ferromagnet: Evidence for a $\pi$ junction},}\
  }\href {\doibase 10.1103/PhysRevLett.86.2427} {\bibfield  {journal} {\bibinfo
   {journal} {Phys. Rev. Lett.}\ }\textbf {\bibinfo {volume} {86}},\ \bibinfo
  {pages} {2427} (\bibinfo {year} {2001})}\BibitemShut {NoStop}%
\bibitem [{\citenamefont {Kontos}\ \emph {et~al.}(2002)\citenamefont {Kontos},
  \citenamefont {Aprili}, \citenamefont {Lesueur}, \citenamefont {Genêt},
  \citenamefont {Stephanidis},\ and\ \citenamefont
  {Boursier}}]{kontos_josephson_2002}%
  \BibitemOpen
  \bibfield  {author} {\bibinfo {author} {\bibfnamefont {T.}~\bibnamefont
  {Kontos}}, \bibinfo {author} {\bibfnamefont {M.}~\bibnamefont {Aprili}},
  \bibinfo {author} {\bibfnamefont {J.}~\bibnamefont {Lesueur}}, \bibinfo
  {author} {\bibfnamefont {F.}~\bibnamefont {Genêt}}, \bibinfo {author}
  {\bibfnamefont {B.}~\bibnamefont {Stephanidis}}, \ and\ \bibinfo {author}
  {\bibfnamefont {R.}~\bibnamefont {Boursier}},\ }\bibfield  {title} {\emph
  {\bibinfo {title} {Josephson {Junction} through a {Thin} {Ferromagnetic}
  {Layer}: {Negative} {Coupling}},}\ }\href {\doibase
  10.1103/PhysRevLett.89.137007} {\bibfield  {journal} {\bibinfo  {journal}
  {Physical Review Letters}\ }\textbf {\bibinfo {volume} {89}},\ \bibinfo
  {pages} {137007} (\bibinfo {year} {2002})}\BibitemShut {NoStop}%
\bibitem [{\citenamefont {Robinson}\ \emph {et~al.}(2006)\citenamefont
  {Robinson}, \citenamefont {Piano}, \citenamefont {Burnell}, \citenamefont
  {Bell},\ and\ \citenamefont {Blamire}}]{robinson_critical_2006}%
  \BibitemOpen
  \bibfield  {author} {\bibinfo {author} {\bibfnamefont {J.~W.~A.}\
  \bibnamefont {Robinson}}, \bibinfo {author} {\bibfnamefont {S.}~\bibnamefont
  {Piano}}, \bibinfo {author} {\bibfnamefont {G.}~\bibnamefont {Burnell}},
  \bibinfo {author} {\bibfnamefont {C.}~\bibnamefont {Bell}}, \ and\ \bibinfo
  {author} {\bibfnamefont {M.~G.}\ \bibnamefont {Blamire}},\ }\bibfield
  {title} {\emph {\bibinfo {title} {Critical {Current} {Oscillations} in
  {Strong} {Ferromagnetic} $\pi$ {Junctions}},}\ }\href {\doibase
  10.1103/PhysRevLett.97.177003} {\bibfield  {journal} {\bibinfo  {journal}
  {Physical Review Letters}\ }\textbf {\bibinfo {volume} {97}},\ \bibinfo
  {pages} {177003} (\bibinfo {year} {2006})}\BibitemShut {NoStop}%
\bibitem [{\citenamefont {Shimizu}\ \emph {et~al.}(1998)\citenamefont
  {Shimizu}, \citenamefont {Horii}, \citenamefont {Takane},\ and\ \citenamefont
  {Isawa}}]{shimizu_multilevel_1998}%
  \BibitemOpen
  \bibfield  {author} {\bibinfo {author} {\bibfnamefont {Y.}~\bibnamefont
  {Shimizu}}, \bibinfo {author} {\bibfnamefont {H.}~\bibnamefont {Horii}},
  \bibinfo {author} {\bibfnamefont {Y.}~\bibnamefont {Takane}}, \ and\ \bibinfo
  {author} {\bibfnamefont {Y.}~\bibnamefont {Isawa}},\ }\bibfield  {title}
  {\emph {\bibinfo {title} {Multilevel {Effect} on the {Josephson} {Current}
  through a {Quantum} {Dot}},}\ }\href {\doibase 10.1143/JPSJ.67.1525}
  {\bibfield  {journal} {\bibinfo  {journal} {Journal of the Physical Society
  of Japan}\ }\textbf {\bibinfo {volume} {67}},\ \bibinfo {pages} {1525}
  (\bibinfo {year} {1998})}\BibitemShut {NoStop}%
\bibitem [{\citenamefont {van Dam}\ \emph {et~al.}(2006)\citenamefont {van
  Dam}, \citenamefont {Nazarov}, \citenamefont {Bakkers}, \citenamefont
  {De~Franceschi},\ and\ \citenamefont {Kouwenhoven}}]{dam_supercurrent_2006}%
  \BibitemOpen
  \bibfield  {author} {\bibinfo {author} {\bibfnamefont {J.~A.}\ \bibnamefont
  {van Dam}}, \bibinfo {author} {\bibfnamefont {Y.~V.}\ \bibnamefont
  {Nazarov}}, \bibinfo {author} {\bibfnamefont {E.~P. A.~M.}\ \bibnamefont
  {Bakkers}}, \bibinfo {author} {\bibfnamefont {S.}~\bibnamefont
  {De~Franceschi}}, \ and\ \bibinfo {author} {\bibfnamefont {L.~P.}\
  \bibnamefont {Kouwenhoven}},\ }\bibfield  {title} {{\selectlanguage
  {english}\emph {\bibinfo {title} {Supercurrent reversal in quantum dots},}\
  }}\href {\doibase 10.1038/nature05018} {\bibfield  {journal} {\bibinfo
  {journal} {Nature}\ }\textbf {\bibinfo {volume} {442}},\ \bibinfo {pages}
  {667} (\bibinfo {year} {2006})}\BibitemShut {NoStop}%
\bibitem [{\citenamefont {Estrada~Saldaña}\ \emph {et~al.}(2020)\citenamefont
  {Estrada~Saldaña}, \citenamefont {Vekris}, \citenamefont {Žitko},
  \citenamefont {Steffensen}, \citenamefont {Krogstrup}, \citenamefont
  {Paaske}, \citenamefont {Grove-Rasmussen},\ and\ \citenamefont
  {Nygård}}]{saldana_two-impurity_2020}%
  \BibitemOpen
  \bibfield  {author} {\bibinfo {author} {\bibfnamefont {J.~C.}\ \bibnamefont
  {Estrada~Saldaña}}, \bibinfo {author} {\bibfnamefont {A.}~\bibnamefont
  {Vekris}}, \bibinfo {author} {\bibfnamefont {R.}~\bibnamefont {Žitko}},
  \bibinfo {author} {\bibfnamefont {G.}~\bibnamefont {Steffensen}}, \bibinfo
  {author} {\bibfnamefont {P.}~\bibnamefont {Krogstrup}}, \bibinfo {author}
  {\bibfnamefont {J.}~\bibnamefont {Paaske}}, \bibinfo {author} {\bibfnamefont
  {K.}~\bibnamefont {Grove-Rasmussen}}, \ and\ \bibinfo {author} {\bibfnamefont
  {J.}~\bibnamefont {Nygård}},\ }\bibfield  {title} {\emph {\bibinfo {title}
  {Two-impurity {Yu}-{Shiba}-{Rusinov} states in coupled quantum dots},}\
  }\href {\doibase 10.1103/PhysRevB.102.195143} {\bibfield  {journal} {\bibinfo
   {journal} {Physical Review B}\ }\textbf {\bibinfo {volume} {102}},\ \bibinfo
  {pages} {195143} (\bibinfo {year} {2020})}\BibitemShut {NoStop}%
\bibitem [{\citenamefont {Mart{\'i}n-Rodero}\ and\ \citenamefont
  {Yeyati}(2011)}]{martin-rodero_josephson_2011}%
  \BibitemOpen
  \bibfield  {author} {\bibinfo {author} {\bibfnamefont {A.}~\bibnamefont
  {Mart{\'i}n-Rodero}}\ and\ \bibinfo {author} {\bibfnamefont {A.~L.}\
  \bibnamefont {Yeyati}},\ }\bibfield  {title} {\emph {\bibinfo {title}
  {Josephson and {Andreev} transport through quantum dots},}\ }\href {\doibase
  10.1080/00018732.2011.624266} {\bibfield  {journal} {\bibinfo  {journal}
  {Advances in Physics}\ }\textbf {\bibinfo {volume} {60}},\ \bibinfo {pages}
  {899} (\bibinfo {year} {2011})}\BibitemShut {NoStop}%
\bibitem [{\citenamefont {De~Franceschi}\ \emph {et~al.}(2010)\citenamefont
  {De~Franceschi}, \citenamefont {Kouwenhoven}, \citenamefont
  {Schönenberger},\ and\ \citenamefont
  {Wernsdorfer}}]{franceschi_hybrid_2010}%
  \BibitemOpen
  \bibfield  {author} {\bibinfo {author} {\bibfnamefont {S.}~\bibnamefont
  {De~Franceschi}}, \bibinfo {author} {\bibfnamefont {L.}~\bibnamefont
  {Kouwenhoven}}, \bibinfo {author} {\bibfnamefont {C.}~\bibnamefont
  {Schönenberger}}, \ and\ \bibinfo {author} {\bibfnamefont {W.}~\bibnamefont
  {Wernsdorfer}},\ }\bibfield  {title} {{\selectlanguage {english}\emph
  {\bibinfo {title} {Hybrid superconductor–quantum dot devices},}\ }}\href
  {\doibase 10.1038/nnano.2010.173} {\bibfield  {journal} {\bibinfo  {journal}
  {Nature Nanotechnology}\ }\textbf {\bibinfo {volume} {5}},\ \bibinfo {pages}
  {703} (\bibinfo {year} {2010})}\BibitemShut {NoStop}%
\bibitem [{\citenamefont {Yu}(1965)}]{yu_bound_1965}%
  \BibitemOpen
  \bibfield  {author} {\bibinfo {author} {\bibfnamefont {L.}~\bibnamefont
  {Yu}},\ }\bibfield  {title} {{\selectlanguage {english}\emph {\bibinfo
  {title} {Bound state in superconductors with paramagnetic impurities},}\
  }}\href {http://wulixb.iphy.ac.cn/EN/abstract/abstract851.shtml} {\bibfield
  {journal} {\bibinfo  {journal} {Acta Phys.Sin.}\ }\textbf {\bibinfo {volume}
  {21}},\ \bibinfo {pages} {75} (\bibinfo {year} {1965})}\BibitemShut {NoStop}%
\bibitem [{\citenamefont {Shiba}(1968)}]{shiba_classical_1968}%
  \BibitemOpen
  \bibfield  {author} {\bibinfo {author} {\bibfnamefont {H.}~\bibnamefont
  {Shiba}},\ }\bibfield  {title} {{\selectlanguage {english}\emph {\bibinfo
  {title} {Classical spins in superconductors},}\ }}\href {\doibase
  10.1143/PTP.40.435} {\bibfield  {journal} {\bibinfo  {journal} {Prog. Theor.
  Phys.}\ }\textbf {\bibinfo {volume} {40}},\ \bibinfo {pages} {435} (\bibinfo
  {year} {1968})}\BibitemShut {NoStop}%
\bibitem [{\citenamefont {Rusinov}(1969)}]{rusinov_superconductivity_1969}%
  \BibitemOpen
  \bibfield  {author} {\bibinfo {author} {\bibfnamefont {A.~I.}\ \bibnamefont
  {Rusinov}},\ }\bibfield  {title} {\emph {\bibinfo {title} {Superconductivity
  near a paramagmetic impurity},}\ }\href@noop {} {\bibfield  {journal}
  {\bibinfo  {journal} {JETP Letters}\ }\textbf {\bibinfo {volume} {9}},\
  \bibinfo {pages} {85} (\bibinfo {year} {1969})}\BibitemShut {NoStop}%
\bibitem [{\citenamefont {Salkola}\ \emph {et~al.}(1997)\citenamefont
  {Salkola}, \citenamefont {Balatsky},\ and\ \citenamefont
  {Schrieffer}}]{salkola_spectral_1997}%
  \BibitemOpen
  \bibfield  {author} {\bibinfo {author} {\bibfnamefont {M.~I.}\ \bibnamefont
  {Salkola}}, \bibinfo {author} {\bibfnamefont {A.~V.}\ \bibnamefont
  {Balatsky}}, \ and\ \bibinfo {author} {\bibfnamefont {J.~R.}\ \bibnamefont
  {Schrieffer}},\ }\bibfield  {title} {\emph {\bibinfo {title} {Spectral
  properties of quasiparticle excitations induced by magnetic moments in
  superconductors},}\ }\href {\doibase 10.1103/PhysRevB.55.12648} {\bibfield
  {journal} {\bibinfo  {journal} {Phys. Rev. B}\ }\textbf {\bibinfo {volume}
  {55}},\ \bibinfo {pages} {12648} (\bibinfo {year} {1997})}\BibitemShut
  {NoStop}%
\bibitem [{\citenamefont {Flatt{\'e}}\ and\ \citenamefont
  {Byers}(1997{\natexlab{a}})}]{flatte_local_1997}%
  \BibitemOpen
  \bibfield  {author} {\bibinfo {author} {\bibfnamefont {M.~E.}\ \bibnamefont
  {Flatt{\'e}}}\ and\ \bibinfo {author} {\bibfnamefont {J.~M.}\ \bibnamefont
  {Byers}},\ }\bibfield  {title} {\emph {\bibinfo {title} {Local electronic
  structure of defects in superconductors},}\ }\href {\doibase
  10.1103/PhysRevB.56.11213} {\bibfield  {journal} {\bibinfo  {journal} {Phys.
  Rev. B}\ }\textbf {\bibinfo {volume} {56}},\ \bibinfo {pages} {11213}
  (\bibinfo {year} {1997}{\natexlab{a}})}\BibitemShut {NoStop}%
\bibitem [{\citenamefont {Flatt{\'e}}\ and\ \citenamefont
  {Byers}(1997{\natexlab{b}})}]{flatte_local_1997-1}%
  \BibitemOpen
  \bibfield  {author} {\bibinfo {author} {\bibfnamefont {M.~E.}\ \bibnamefont
  {Flatt{\'e}}}\ and\ \bibinfo {author} {\bibfnamefont {J.~M.}\ \bibnamefont
  {Byers}},\ }\bibfield  {title} {\emph {\bibinfo {title} {Local electronic
  structure of a single magnetic impurity in a superconductor},}\ }\href
  {\doibase 10.1103/PhysRevLett.78.3761} {\bibfield  {journal} {\bibinfo
  {journal} {Phys. Rev. Lett.}\ }\textbf {\bibinfo {volume} {78}},\ \bibinfo
  {pages} {3761} (\bibinfo {year} {1997}{\natexlab{b}})}\BibitemShut {NoStop}%
\bibitem [{\citenamefont {Balatsky}\ \emph {et~al.}(2006)\citenamefont
  {Balatsky}, \citenamefont {Vekhter},\ and\ \citenamefont
  {Zhu}}]{balatsky_impurity-induced_2006}%
  \BibitemOpen
  \bibfield  {author} {\bibinfo {author} {\bibfnamefont {A.~V.}\ \bibnamefont
  {Balatsky}}, \bibinfo {author} {\bibfnamefont {I.}~\bibnamefont {Vekhter}}, \
  and\ \bibinfo {author} {\bibfnamefont {J.-X.}\ \bibnamefont {Zhu}},\
  }\bibfield  {title} {\emph {\bibinfo {title} {Impurity-induced states in
  conventional and unconventional superconductors},}\ }\href {\doibase
  10.1103/RevModPhys.78.373} {\bibfield  {journal} {\bibinfo  {journal} {Rev.
  Mod. Phys.}\ }\textbf {\bibinfo {volume} {78}},\ \bibinfo {pages} {373}
  (\bibinfo {year} {2006})}\BibitemShut {NoStop}%
\bibitem [{\citenamefont {Farinacci}\ \emph {et~al.}(2018)\citenamefont
  {Farinacci}, \citenamefont {Ahmadi}, \citenamefont {Reecht}, \citenamefont
  {Ruby}, \citenamefont {Bogdanoff}, \citenamefont {Peters}, \citenamefont
  {Heinrich}, \citenamefont {von Oppen},\ and\ \citenamefont
  {Franke}}]{farinacci_tuning_2018}%
  \BibitemOpen
  \bibfield  {author} {\bibinfo {author} {\bibfnamefont {L.}~\bibnamefont
  {Farinacci}}, \bibinfo {author} {\bibfnamefont {G.}~\bibnamefont {Ahmadi}},
  \bibinfo {author} {\bibfnamefont {G.}~\bibnamefont {Reecht}}, \bibinfo
  {author} {\bibfnamefont {M.}~\bibnamefont {Ruby}}, \bibinfo {author}
  {\bibfnamefont {N.}~\bibnamefont {Bogdanoff}}, \bibinfo {author}
  {\bibfnamefont {O.}~\bibnamefont {Peters}}, \bibinfo {author} {\bibfnamefont
  {B.~W.}\ \bibnamefont {Heinrich}}, \bibinfo {author} {\bibfnamefont
  {F.}~\bibnamefont {von Oppen}}, \ and\ \bibinfo {author} {\bibfnamefont
  {K.~J.}\ \bibnamefont {Franke}},\ }\bibfield  {title} {\emph {\bibinfo
  {title} {Tuning the {Coupling} of an {Individual} {Magnetic} {Impurity} to a
  {Superconductor}: {Quantum} {Phase} {Transition} and {Transport}},}\ }\href
  {\doibase 10.1103/PhysRevLett.121.196803} {\bibfield  {journal} {\bibinfo
  {journal} {Physical Review Letters}\ }\textbf {\bibinfo {volume} {121}},\
  \bibinfo {pages} {196803} (\bibinfo {year} {2018})}\BibitemShut {NoStop}%
\bibitem [{\citenamefont {Malavolti}\ \emph {et~al.}(2018)\citenamefont
  {Malavolti}, \citenamefont {Briganti}, \citenamefont {H{\"a}nze},
  \citenamefont {Serrano}, \citenamefont {Cimatti}, \citenamefont {McMurtrie},
  \citenamefont {Otero}, \citenamefont {Ohresser}, \citenamefont {Totti},
  \citenamefont {Mannini}, \citenamefont {Sessoli},\ and\ \citenamefont
  {Loth}}]{malavolti_tunable_2018}%
  \BibitemOpen
  \bibfield  {author} {\bibinfo {author} {\bibfnamefont {L.}~\bibnamefont
  {Malavolti}}, \bibinfo {author} {\bibfnamefont {M.}~\bibnamefont {Briganti}},
  \bibinfo {author} {\bibfnamefont {M.}~\bibnamefont {H{\"a}nze}}, \bibinfo
  {author} {\bibfnamefont {G.}~\bibnamefont {Serrano}}, \bibinfo {author}
  {\bibfnamefont {I.}~\bibnamefont {Cimatti}}, \bibinfo {author} {\bibfnamefont
  {G.}~\bibnamefont {McMurtrie}}, \bibinfo {author} {\bibfnamefont
  {E.}~\bibnamefont {Otero}}, \bibinfo {author} {\bibfnamefont
  {P.}~\bibnamefont {Ohresser}}, \bibinfo {author} {\bibfnamefont
  {F.}~\bibnamefont {Totti}}, \bibinfo {author} {\bibfnamefont
  {M.}~\bibnamefont {Mannini}}, \bibinfo {author} {\bibfnamefont
  {R.}~\bibnamefont {Sessoli}}, \ and\ \bibinfo {author} {\bibfnamefont
  {S.}~\bibnamefont {Loth}},\ }\bibfield  {title} {\emph {\bibinfo {title}
  {Tunable {Spin}-{Superconductor} {Coupling} of {Spin} 1/2 {Vanadyl}
  {Phthalocyanine} {Molecules}},}\ }\href {\doibase
  10.1021/acs.nanolett.8b03921} {\bibfield  {journal} {\bibinfo  {journal}
  {Nano Letters}\ }\textbf {\bibinfo {volume} {18}},\ \bibinfo {pages} {7955}
  (\bibinfo {year} {2018})}\BibitemShut {NoStop}%
\bibitem [{\citenamefont {Franke}\ \emph {et~al.}(2011)\citenamefont {Franke},
  \citenamefont {Schulze},\ and\ \citenamefont
  {Pascual}}]{franke_competition_2011}%
  \BibitemOpen
  \bibfield  {author} {\bibinfo {author} {\bibfnamefont {K.~J.}\ \bibnamefont
  {Franke}}, \bibinfo {author} {\bibfnamefont {G.}~\bibnamefont {Schulze}}, \
  and\ \bibinfo {author} {\bibfnamefont {J.~I.}\ \bibnamefont {Pascual}},\
  }\bibfield  {title} {\emph {\bibinfo {title} {Competition of superconducting
  phenomena and {Kondo} screening at the nanoscale},}\ }\href {\doibase
  10.1126/science.1202204} {\bibfield  {journal} {\bibinfo  {journal}
  {Science}\ }\textbf {\bibinfo {volume} {332}},\ \bibinfo {pages} {940 }
  (\bibinfo {year} {2011})}\BibitemShut {NoStop}%
\bibitem [{\citenamefont {Bauer}\ \emph {et~al.}(2013)\citenamefont {Bauer},
  \citenamefont {Pascual},\ and\ \citenamefont
  {Franke}}]{bauer_microscopic_2013}%
  \BibitemOpen
  \bibfield  {author} {\bibinfo {author} {\bibfnamefont {J.}~\bibnamefont
  {Bauer}}, \bibinfo {author} {\bibfnamefont {J.~I.}\ \bibnamefont {Pascual}},
  \ and\ \bibinfo {author} {\bibfnamefont {K.~J.}\ \bibnamefont {Franke}},\
  }\bibfield  {title} {\emph {\bibinfo {title} {Microscopic resolution of the
  interplay of kondo screening and superconducting pairing: Mn-phthalocyanine
  molecules adsorbed on superconducting pb(111)},}\ }\href {\doibase
  10.1103/PhysRevB.87.075125} {\bibfield  {journal} {\bibinfo  {journal} {Phys.
  Rev. B}\ }\textbf {\bibinfo {volume} {87}},\ \bibinfo {pages} {075125}
  (\bibinfo {year} {2013})}\BibitemShut {NoStop}%
\bibitem [{\citenamefont {Kamlapure}\ \emph {et~al.}(2019)\citenamefont
  {Kamlapure}, \citenamefont {Cornils}, \citenamefont {Žitko}, \citenamefont
  {Valentyuk}, \citenamefont {Mozara}, \citenamefont {Pradhan}, \citenamefont
  {Fransson}, \citenamefont {Lichtenstein}, \citenamefont {Wiebe},\ and\
  \citenamefont {Wiesendanger}}]{kamlapure_investigation_2019}%
  \BibitemOpen
  \bibfield  {author} {\bibinfo {author} {\bibfnamefont {A.}~\bibnamefont
  {Kamlapure}}, \bibinfo {author} {\bibfnamefont {L.}~\bibnamefont {Cornils}},
  \bibinfo {author} {\bibfnamefont {R.}~\bibnamefont {Žitko}}, \bibinfo
  {author} {\bibfnamefont {M.}~\bibnamefont {Valentyuk}}, \bibinfo {author}
  {\bibfnamefont {R.}~\bibnamefont {Mozara}}, \bibinfo {author} {\bibfnamefont
  {S.}~\bibnamefont {Pradhan}}, \bibinfo {author} {\bibfnamefont
  {J.}~\bibnamefont {Fransson}}, \bibinfo {author} {\bibfnamefont {A.~I.}\
  \bibnamefont {Lichtenstein}}, \bibinfo {author} {\bibfnamefont
  {J.}~\bibnamefont {Wiebe}}, \ and\ \bibinfo {author} {\bibfnamefont
  {R.}~\bibnamefont {Wiesendanger}},\ }\bibfield  {title} {\emph {\bibinfo
  {title} {Investigation of the {Yu}-{Shiba}-{Rusinov} states of a
  multi-impurity {Kondo} system},}\ }\href {http://arxiv.org/abs/1911.03794}
  {\bibfield  {journal} {\bibinfo  {journal} {arXiv:1911.03794 [cond-mat]}\ }
  (\bibinfo {year} {2019})}\BibitemShut {NoStop}%
\bibitem [{\citenamefont {Huang}\ \emph {et~al.}(2020)\citenamefont {Huang},
  \citenamefont {Drost}, \citenamefont {Senkpiel}, \citenamefont {Padurariu},
  \citenamefont {Kubala}, \citenamefont {Yeyati}, \citenamefont {Cuevas},
  \citenamefont {Ankerhold}, \citenamefont {Kern},\ and\ \citenamefont
  {Ast}}]{huang_quantum_2020}%
  \BibitemOpen
  \bibfield  {author} {\bibinfo {author} {\bibfnamefont {H.}~\bibnamefont
  {Huang}}, \bibinfo {author} {\bibfnamefont {R.}~\bibnamefont {Drost}},
  \bibinfo {author} {\bibfnamefont {J.}~\bibnamefont {Senkpiel}}, \bibinfo
  {author} {\bibfnamefont {C.}~\bibnamefont {Padurariu}}, \bibinfo {author}
  {\bibfnamefont {B.}~\bibnamefont {Kubala}}, \bibinfo {author} {\bibfnamefont
  {A.~L.}\ \bibnamefont {Yeyati}}, \bibinfo {author} {\bibfnamefont {J.~C.}\
  \bibnamefont {Cuevas}}, \bibinfo {author} {\bibfnamefont {J.}~\bibnamefont
  {Ankerhold}}, \bibinfo {author} {\bibfnamefont {K.}~\bibnamefont {Kern}}, \
  and\ \bibinfo {author} {\bibfnamefont {C.~R.}\ \bibnamefont {Ast}},\
  }\bibfield  {title} {{\selectlanguage {english}\emph {\bibinfo {title}
  {Quantum phase transitions and the role of impurity-substrate hybridization
  in {Yu}-{Shiba}-{Rusinov} states},}\ }}\href {\doibase
  10.1038/s42005-020-00469-0} {\bibfield  {journal} {\bibinfo  {journal}
  {Communications Physics}\ }\textbf {\bibinfo {volume} {3}},\ \bibinfo {pages}
  {199} (\bibinfo {year} {2020})}\BibitemShut {NoStop}%
\bibitem [{\citenamefont {Kezilebieke}\ \emph {et~al.}(2019)\citenamefont
  {Kezilebieke}, \citenamefont {Žitko}, \citenamefont {Dvorak}, \citenamefont
  {Ojanen},\ and\ \citenamefont {Liljeroth}}]{kezilebieke_observation_2019}%
  \BibitemOpen
  \bibfield  {author} {\bibinfo {author} {\bibfnamefont {S.}~\bibnamefont
  {Kezilebieke}}, \bibinfo {author} {\bibfnamefont {R.}~\bibnamefont {Žitko}},
  \bibinfo {author} {\bibfnamefont {M.}~\bibnamefont {Dvorak}}, \bibinfo
  {author} {\bibfnamefont {T.}~\bibnamefont {Ojanen}}, \ and\ \bibinfo {author}
  {\bibfnamefont {P.}~\bibnamefont {Liljeroth}},\ }\bibfield  {title} {\emph
  {\bibinfo {title} {Observation of {Coexistence} of {Yu}-{Shiba}-{Rusinov}
  {States} and {Spin}-{Flip} {Excitations}},}\ }\href {\doibase
  10.1021/acs.nanolett.9b01583} {\bibfield  {journal} {\bibinfo  {journal}
  {Nano Letters}\ }\textbf {\bibinfo {volume} {19}},\ \bibinfo {pages} {4614}
  (\bibinfo {year} {2019})}\BibitemShut {NoStop}%
\bibitem [{\citenamefont {Ternes}\ \emph {et~al.}(2011)\citenamefont {Ternes},
  \citenamefont {Gonz\'{a}lez}, \citenamefont {Lutz}, \citenamefont {Hapala},
  \citenamefont {Giessibl}, \citenamefont {Jel\'{i}nek},\ and\ \citenamefont
  {Heinrich}}]{ternes_interplay_2011}%
  \BibitemOpen
  \bibfield  {author} {\bibinfo {author} {\bibfnamefont {M.}~\bibnamefont
  {Ternes}}, \bibinfo {author} {\bibfnamefont {C.}~\bibnamefont
  {Gonz\'{a}lez}}, \bibinfo {author} {\bibfnamefont {C.~P.}\ \bibnamefont
  {Lutz}}, \bibinfo {author} {\bibfnamefont {P.}~\bibnamefont {Hapala}},
  \bibinfo {author} {\bibfnamefont {F.~J.}\ \bibnamefont {Giessibl}}, \bibinfo
  {author} {\bibfnamefont {P.}~\bibnamefont {Jel\'{i}nek}}, \ and\ \bibinfo
  {author} {\bibfnamefont {A.~J.}\ \bibnamefont {Heinrich}},\ }\bibfield
  {title} {\emph {\bibinfo {title} {Interplay of conductance, force, and
  structural change in metallic point contacts},}\ }\href {\doibase
  10.1103/PhysRevLett.106.016802} {\bibfield  {journal} {\bibinfo  {journal}
  {Physical Review Letters}\ }\textbf {\bibinfo {volume} {106}},\ \bibinfo
  {pages} {016802} (\bibinfo {year} {2011})}\BibitemShut {NoStop}%
\bibitem [{\citenamefont {Ast}\ \emph {et~al.}(2016)\citenamefont {Ast},
  \citenamefont {J{\"a}ck}, \citenamefont {Senkpiel}, \citenamefont {Eltschka},
  \citenamefont {Etzkorn}, \citenamefont {Ankerhold},\ and\ \citenamefont
  {Kern}}]{ast_sensing_2016}%
  \BibitemOpen
  \bibfield  {author} {\bibinfo {author} {\bibfnamefont {C.~R.}\ \bibnamefont
  {Ast}}, \bibinfo {author} {\bibfnamefont {B.}~\bibnamefont {J{\"a}ck}},
  \bibinfo {author} {\bibfnamefont {J.}~\bibnamefont {Senkpiel}}, \bibinfo
  {author} {\bibfnamefont {M.}~\bibnamefont {Eltschka}}, \bibinfo {author}
  {\bibfnamefont {M.}~\bibnamefont {Etzkorn}}, \bibinfo {author} {\bibfnamefont
  {J.}~\bibnamefont {Ankerhold}}, \ and\ \bibinfo {author} {\bibfnamefont
  {K.}~\bibnamefont {Kern}},\ }\bibfield  {title} {\emph {\bibinfo {title}
  {Sensing the quantum limit in scanning tunnelling spectroscopy},}\ }\href
  {\doibase 10.1038/ncomms13009} {\bibfield  {journal} {\bibinfo  {journal}
  {Nature Communications}\ }\textbf {\bibinfo {volume} {7}},\ \bibinfo {pages}
  {13009} (\bibinfo {year} {2016})}\BibitemShut {NoStop}%
\bibitem [{\citenamefont {Senkpiel}\ \emph {et~al.}(2020)\citenamefont
  {Senkpiel}, \citenamefont {Dambach}, \citenamefont {Etzkorn}, \citenamefont
  {Drost}, \citenamefont {Padurariu}, \citenamefont {Kubala}, \citenamefont
  {Belzig}, \citenamefont {Yeyati}, \citenamefont {Cuevas}, \citenamefont
  {Ankerhold}, \citenamefont {Ast},\ and\ \citenamefont
  {Kern}}]{senkpiel_single_2020}%
  \BibitemOpen
  \bibfield  {author} {\bibinfo {author} {\bibfnamefont {J.}~\bibnamefont
  {Senkpiel}}, \bibinfo {author} {\bibfnamefont {S.}~\bibnamefont {Dambach}},
  \bibinfo {author} {\bibfnamefont {M.}~\bibnamefont {Etzkorn}}, \bibinfo
  {author} {\bibfnamefont {R.}~\bibnamefont {Drost}}, \bibinfo {author}
  {\bibfnamefont {C.}~\bibnamefont {Padurariu}}, \bibinfo {author}
  {\bibfnamefont {B.}~\bibnamefont {Kubala}}, \bibinfo {author} {\bibfnamefont
  {W.}~\bibnamefont {Belzig}}, \bibinfo {author} {\bibfnamefont {A.~L.}\
  \bibnamefont {Yeyati}}, \bibinfo {author} {\bibfnamefont {J.~C.}\
  \bibnamefont {Cuevas}}, \bibinfo {author} {\bibfnamefont {J.}~\bibnamefont
  {Ankerhold}}, \bibinfo {author} {\bibfnamefont {C.~R.}\ \bibnamefont {Ast}},
  \ and\ \bibinfo {author} {\bibfnamefont {K.}~\bibnamefont {Kern}},\
  }\bibfield  {title} {{\selectlanguage {english}\emph {\bibinfo {title}
  {Single channel {Josephson} effect in a high transmission atomic contact},}\
  }}\href {\doibase 10.1038/s42005-020-00397-z} {\bibfield  {journal} {\bibinfo
   {journal} {Communications Physics}\ }\textbf {\bibinfo {volume} {3}},\
  \bibinfo {pages} {131} (\bibinfo {year} {2020})}\BibitemShut {NoStop}%
\bibitem [{\citenamefont {Ingold}\ \emph {et~al.}(1994)\citenamefont {Ingold},
  \citenamefont {Grabert},\ and\ \citenamefont
  {Eberhardt}}]{ingold_cooper-pair_1994}%
  \BibitemOpen
  \bibfield  {author} {\bibinfo {author} {\bibfnamefont {G.}~\bibnamefont
  {Ingold}}, \bibinfo {author} {\bibfnamefont {H.}~\bibnamefont {Grabert}}, \
  and\ \bibinfo {author} {\bibfnamefont {U.}~\bibnamefont {Eberhardt}},\
  }\bibfield  {title} {\emph {\bibinfo {title} {Cooper-pair current through
  ultrasmall josephson junctions},}\ }\href {\doibase 10.1103/PhysRevB.50.395}
  {\bibfield  {journal} {\bibinfo  {journal} {Physical Review B}\ }\textbf
  {\bibinfo {volume} {50}},\ \bibinfo {pages} {395} (\bibinfo {year}
  {1994})}\BibitemShut {NoStop}%
\bibitem [{\citenamefont {Devoret}\ \emph {et~al.}(1990)\citenamefont
  {Devoret}, \citenamefont {Esteve}, \citenamefont {Grabert}, \citenamefont
  {Ingold}, \citenamefont {Pothier},\ and\ \citenamefont
  {Urbina}}]{devoret_effect_1990}%
  \BibitemOpen
  \bibfield  {author} {\bibinfo {author} {\bibfnamefont {M.~H.}\ \bibnamefont
  {Devoret}}, \bibinfo {author} {\bibfnamefont {D.}~\bibnamefont {Esteve}},
  \bibinfo {author} {\bibfnamefont {H.}~\bibnamefont {Grabert}}, \bibinfo
  {author} {\bibfnamefont {G.}~\bibnamefont {Ingold}}, \bibinfo {author}
  {\bibfnamefont {H.}~\bibnamefont {Pothier}}, \ and\ \bibinfo {author}
  {\bibfnamefont {C.}~\bibnamefont {Urbina}},\ }\bibfield  {title} {\emph
  {\bibinfo {title} {Effect of the electromagnetic environment on the coulomb
  blockade in ultrasmall tunnel junctions},}\ }\href {\doibase
  10.1103/PhysRevLett.64.1824} {\bibfield  {journal} {\bibinfo  {journal}
  {Physical Review Letters}\ }\textbf {\bibinfo {volume} {64}},\ \bibinfo
  {pages} {1824} (\bibinfo {year} {1990})}\BibitemShut {NoStop}%
\bibitem [{\citenamefont {Averin}\ \emph {et~al.}(1990)\citenamefont {Averin},
  \citenamefont {Nazarov},\ and\ \citenamefont
  {Odintsov}}]{averin_incoherent_1990}%
  \BibitemOpen
  \bibfield  {author} {\bibinfo {author} {\bibfnamefont {D.}~\bibnamefont
  {Averin}}, \bibinfo {author} {\bibfnamefont {Y.}~\bibnamefont {Nazarov}}, \
  and\ \bibinfo {author} {\bibfnamefont {A.}~\bibnamefont {Odintsov}},\
  }\bibfield  {title} {\emph {\bibinfo {title} {Incoherent tunneling of the
  cooper pairs and magnetic flux quanta in ultrasmall josephson junctions},}\
  }\href {\doibase http://dx.doi.org/10.1016/S0921-4526(09)80058-6} {\bibfield
  {journal} {\bibinfo  {journal} {Physica B: Condensed Matter}\ }\textbf
  {\bibinfo {volume} {165-166}},\ \bibinfo {pages} {945} (\bibinfo {year}
  {1990})}\BibitemShut {NoStop}%
\bibitem [{gap()}]{gapnote}%
  \BibitemOpen
  \href@noop {} {}\bibinfo {note} {We actually find a small gap of about
  $11\,\upmu$eV at the YSR energy minimum, which is needed to properly fit the
  experimental data to the theoretical model. The existence of a small gap has
  actually being predicted in the context of a mean-field theory where the
  local variation of the order parameter close to the impurity was determined
  selfconsistently \cite{salkola_spectral_1997}.}\BibitemShut {Stop}%
\bibitem [{sup()}]{supinf}%
  \BibitemOpen
  \href@noop {} {}\bibinfo {note} {See Supporting Information}\BibitemShut
  {NoStop}%
\bibitem [{\citenamefont {J\"ack}\ \emph {et~al.}(2016)\citenamefont {J\"ack},
  \citenamefont {Eltschka}, \citenamefont {Assig}, \citenamefont {Etzkorn},
  \citenamefont {Ast},\ and\ \citenamefont {Kern}}]{jack_critical_2016}%
  \BibitemOpen
  \bibfield  {author} {\bibinfo {author} {\bibfnamefont {B.}~\bibnamefont
  {J\"ack}}, \bibinfo {author} {\bibfnamefont {M.}~\bibnamefont {Eltschka}},
  \bibinfo {author} {\bibfnamefont {M.}~\bibnamefont {Assig}}, \bibinfo
  {author} {\bibfnamefont {M.}~\bibnamefont {Etzkorn}}, \bibinfo {author}
  {\bibfnamefont {C.~R.}\ \bibnamefont {Ast}}, \ and\ \bibinfo {author}
  {\bibfnamefont {K.}~\bibnamefont {Kern}},\ }\bibfield  {title} {\emph
  {\bibinfo {title} {Critical {Josephson} current in the dynamical {Coulomb}
  blockade regime},}\ }\href {\doibase 10.1103/PhysRevB.93.020504} {\bibfield
  {journal} {\bibinfo  {journal} {Phys. Rev. B}\ }\textbf {\bibinfo {volume}
  {93}},\ \bibinfo {pages} {020504} (\bibinfo {year} {2016})}\BibitemShut
  {NoStop}%
\bibitem [{\citenamefont {Ambegaokar}\ and\ \citenamefont
  {Baratoff}(1963)}]{ambegaokar_tunneling_1963}%
  \BibitemOpen
  \bibfield  {author} {\bibinfo {author} {\bibfnamefont {V.}~\bibnamefont
  {Ambegaokar}}\ and\ \bibinfo {author} {\bibfnamefont {A.}~\bibnamefont
  {Baratoff}},\ }\bibfield  {title} {\emph {\bibinfo {title} {Tunneling between
  superconductors},}\ }\href {\doibase 10.1103/PhysRevLett.10.486} {\bibfield
  {journal} {\bibinfo  {journal} {Phys. Rev. Lett.}\ }\textbf {\bibinfo
  {volume} {10}},\ \bibinfo {pages} {486} (\bibinfo {year} {1963})}\BibitemShut
  {NoStop}%
\bibitem [{\citenamefont {Spivak}\ and\ \citenamefont
  {Kivelson}(1991)}]{spivak_negative_1991}%
  \BibitemOpen
  \bibfield  {author} {\bibinfo {author} {\bibfnamefont {B.~I.}\ \bibnamefont
  {Spivak}}\ and\ \bibinfo {author} {\bibfnamefont {S.~A.}\ \bibnamefont
  {Kivelson}},\ }\bibfield  {title} {\emph {\bibinfo {title} {Negative local
  superfluid densities: {The} difference between dirty superconductors and
  dirty {Bose} liquids},}\ }\href {\doibase 10.1103/PhysRevB.43.3740}
  {\bibfield  {journal} {\bibinfo  {journal} {Physical Review B}\ }\textbf
  {\bibinfo {volume} {43}},\ \bibinfo {pages} {3740} (\bibinfo {year}
  {1991})}\BibitemShut {NoStop}%
\end{thebibliography}

\begin{thebibliography}{8}%
\makeatletter
\providecommand \@ifxundefined [1]{%
 \@ifx{#1\undefined}
}%
\providecommand \@ifnum [1]{%
 \ifnum #1\expandafter \@firstoftwo
 \else \expandafter \@secondoftwo
 \fi
}%
\providecommand \@ifx [1]{%
 \ifx #1\expandafter \@firstoftwo
 \else \expandafter \@secondoftwo
 \fi
}%
\providecommand \natexlab [1]{#1}%
\providecommand \emph  [1]{``#1''}%
\providecommand \bibnamefont  [1]{#1}%
\providecommand \bibfnamefont [1]{#1}%
\providecommand \citenamefont [1]{#1}%
\providecommand \href@noop [0]{\@secondoftwo}%
\providecommand \href [0]{\begingroup \@sanitize@url \@href}%
\providecommand \@href[1]{\@@startlink{#1}\@@href}%
\providecommand \@@href[1]{\endgroup#1\@@endlink}%
\providecommand \@sanitize@url [0]{\catcode `\\12\catcode `\$12\catcode
  `\&12\catcode `\#12\catcode `\^12\catcode `\_12\catcode `\%12\relax}%
\providecommand \@@startlink[1]{}%
\providecommand \@@endlink[0]{}%
\providecommand \url  [0]{\begingroup\@sanitize@url \@url }%
\providecommand \@url [1]{\endgroup\@href {#1}{\urlprefix }}%
\providecommand \urlprefix  [0]{URL }%
\providecommand \Eprint [0]{\href }%
\providecommand \doibase [0]{http://dx.doi.org/}%
\providecommand \selectlanguage [0]{\@gobble}%
\providecommand \bibinfo  [0]{\@secondoftwo}%
\providecommand \bibfield  [0]{\@secondoftwo}%
\providecommand \translation [1]{[#1]}%
\providecommand \BibitemOpen [0]{}%
\providecommand \bibitemStop [0]{}%
\providecommand \bibitemNoStop [0]{.\EOS\space}%
\providecommand \EOS [0]{\spacefactor3000\relax}%
\providecommand \BibitemShut  [1]{\csname bibitem#1\endcsname}%
\let\auto@bib@innerbib\@empty
\bibitem [{\citenamefont {Koller}\ \emph {et~al.}(2001)\citenamefont {Koller},
  \citenamefont {Bergermayer}, \citenamefont {Kresse}, \citenamefont
  {Hebenstreit}, \citenamefont {Konvicka}, \citenamefont {Schmid},
  \citenamefont {Podloucky},\ and\ \citenamefont
  {Varga}}]{si_koller_structure_2001}%
  \BibitemOpen
  \bibfield  {author} {\bibinfo {author} {\bibfnamefont {R.}~\bibnamefont
  {Koller}}, \bibinfo {author} {\bibfnamefont {W.}~\bibnamefont {Bergermayer}},
  \bibinfo {author} {\bibfnamefont {G.}~\bibnamefont {Kresse}}, \bibinfo
  {author} {\bibfnamefont {E.~L.~D.}\ \bibnamefont {Hebenstreit}}, \bibinfo
  {author} {\bibfnamefont {C.}~\bibnamefont {Konvicka}}, \bibinfo {author}
  {\bibfnamefont {M.}~\bibnamefont {Schmid}}, \bibinfo {author} {\bibfnamefont
  {R.}~\bibnamefont {Podloucky}}, \ and\ \bibinfo {author} {\bibfnamefont
  {P.}~\bibnamefont {Varga}},\ }\bibfield  {title} {\emph {\bibinfo {title}
  {The structure of the oxygen induced (1$\times$5) reconstruction of
  {V}(100)},}\ }\href {\doibase 10.1016/S0039-6028(01)00978-5} {\bibfield
  {journal} {\bibinfo  {journal} {Surface Science}\ }\textbf {\bibinfo {volume}
  {480}},\ \bibinfo {pages} {11} (\bibinfo {year} {2001})}\BibitemShut
  {NoStop}%
\bibitem [{\citenamefont {Kralj}\ \emph {et~al.}(2003)\citenamefont {Kralj},
  \citenamefont {Pervan}, \citenamefont {Milun}, \citenamefont {Wandelt},
  \citenamefont {Mandrino},\ and\ \citenamefont {Jenko}}]{si_kralj_hraes_2003}%
  \BibitemOpen
  \bibfield  {author} {\bibinfo {author} {\bibfnamefont {M.}~\bibnamefont
  {Kralj}}, \bibinfo {author} {\bibfnamefont {P.}~\bibnamefont {Pervan}},
  \bibinfo {author} {\bibfnamefont {M.}~\bibnamefont {Milun}}, \bibinfo
  {author} {\bibfnamefont {K.}~\bibnamefont {Wandelt}}, \bibinfo {author}
  {\bibfnamefont {D.}~\bibnamefont {Mandrino}}, \ and\ \bibinfo {author}
  {\bibfnamefont {M.}~\bibnamefont {Jenko}},\ }\bibfield  {title} {\emph
  {\bibinfo {title} {{HRAES}, {STM} and {ARUPS} study of (5$\times$1)
  reconstructed {V}(100)},}\ }\href
  {https://www.sciencedirect.com/science/article/pii/S003960280202647X}
  {\bibfield  {journal} {\bibinfo  {journal} {Surface Science}\ }\textbf
  {\bibinfo {volume} {526}},\ \bibinfo {pages} {166} (\bibinfo {year}
  {2003})}\BibitemShut {NoStop}%
\bibitem [{\citenamefont {Huang}\ \emph
  {et~al.}(2020{\natexlab{a}})\citenamefont {Huang}, \citenamefont {Padurariu},
  \citenamefont {Senkpiel}, \citenamefont {Drost}, \citenamefont {Yeyati},
  \citenamefont {Cuevas}, \citenamefont {Kubala}, \citenamefont {Ankerhold},
  \citenamefont {Kern},\ and\ \citenamefont {Ast}}]{si_huang_tunnelling_2020}%
  \BibitemOpen
  \bibfield  {author} {\bibinfo {author} {\bibfnamefont {H.}~\bibnamefont
  {Huang}}, \bibinfo {author} {\bibfnamefont {C.}~\bibnamefont {Padurariu}},
  \bibinfo {author} {\bibfnamefont {J.}~\bibnamefont {Senkpiel}}, \bibinfo
  {author} {\bibfnamefont {R.}~\bibnamefont {Drost}}, \bibinfo {author}
  {\bibfnamefont {A.~L.}\ \bibnamefont {Yeyati}}, \bibinfo {author}
  {\bibfnamefont {J.~C.}\ \bibnamefont {Cuevas}}, \bibinfo {author}
  {\bibfnamefont {B.}~\bibnamefont {Kubala}}, \bibinfo {author} {\bibfnamefont
  {J.}~\bibnamefont {Ankerhold}}, \bibinfo {author} {\bibfnamefont
  {K.}~\bibnamefont {Kern}}, \ and\ \bibinfo {author} {\bibfnamefont {C.~R.}\
  \bibnamefont {Ast}},\ }\bibfield  {title} {{\selectlanguage {english}\emph
  {\bibinfo {title} {Tunnelling dynamics between superconducting bound states
  at the atomic limit},}\ }}\href {\doibase 10.1038/s41567-020-0971-0}
  {\bibfield  {journal} {\bibinfo  {journal} {Nature Physics}\ }\textbf
  {\bibinfo {volume} {16}},\ \bibinfo {pages} {1227–1231} (\bibinfo {year}
  {2020}{\natexlab{a}})}\BibitemShut {NoStop}%
\bibitem [{\citenamefont {Senkpiel}\ \emph {et~al.}(2020)\citenamefont
  {Senkpiel}, \citenamefont {Dambach}, \citenamefont {Etzkorn}, \citenamefont
  {Drost}, \citenamefont {Padurariu}, \citenamefont {Kubala}, \citenamefont
  {Belzig}, \citenamefont {Yeyati}, \citenamefont {Cuevas}, \citenamefont
  {Ankerhold}, \citenamefont {Ast},\ and\ \citenamefont
  {Kern}}]{si_senkpiel_single_2020}%
  \BibitemOpen
  \bibfield  {author} {\bibinfo {author} {\bibfnamefont {J.}~\bibnamefont
  {Senkpiel}}, \bibinfo {author} {\bibfnamefont {S.}~\bibnamefont {Dambach}},
  \bibinfo {author} {\bibfnamefont {M.}~\bibnamefont {Etzkorn}}, \bibinfo
  {author} {\bibfnamefont {R.}~\bibnamefont {Drost}}, \bibinfo {author}
  {\bibfnamefont {C.}~\bibnamefont {Padurariu}}, \bibinfo {author}
  {\bibfnamefont {B.}~\bibnamefont {Kubala}}, \bibinfo {author} {\bibfnamefont
  {W.}~\bibnamefont {Belzig}}, \bibinfo {author} {\bibfnamefont {A.~L.}\
  \bibnamefont {Yeyati}}, \bibinfo {author} {\bibfnamefont {J.~C.}\
  \bibnamefont {Cuevas}}, \bibinfo {author} {\bibfnamefont {J.}~\bibnamefont
  {Ankerhold}}, \bibinfo {author} {\bibfnamefont {C.~R.}\ \bibnamefont {Ast}},
  \ and\ \bibinfo {author} {\bibfnamefont {K.}~\bibnamefont {Kern}},\
  }\bibfield  {title} {{\selectlanguage {english}\emph {\bibinfo {title}
  {Single channel {Josephson} effect in a high transmission atomic contact},}\
  }}\href {\doibase 10.1038/s42005-020-00397-z} {\bibfield  {journal} {\bibinfo
   {journal} {Communications Physics}\ }\textbf {\bibinfo {volume} {3}},\
  \bibinfo {pages} {131} (\bibinfo {year} {2020})}\BibitemShut {NoStop}%
\bibitem [{\citenamefont {Villas}\ \emph {et~al.}(2020)\citenamefont {Villas},
  \citenamefont {Klees}, \citenamefont {Huang}, \citenamefont {Ast},
  \citenamefont {Rastelli}, \citenamefont {Belzig},\ and\ \citenamefont
  {Cuevas}}]{si_villas_interplay_2020}%
  \BibitemOpen
  \bibfield  {author} {\bibinfo {author} {\bibfnamefont {A.}~\bibnamefont
  {Villas}}, \bibinfo {author} {\bibfnamefont {R.~L.}\ \bibnamefont {Klees}},
  \bibinfo {author} {\bibfnamefont {H.}~\bibnamefont {Huang}}, \bibinfo
  {author} {\bibfnamefont {C.~R.}\ \bibnamefont {Ast}}, \bibinfo {author}
  {\bibfnamefont {G.}~\bibnamefont {Rastelli}}, \bibinfo {author}
  {\bibfnamefont {W.}~\bibnamefont {Belzig}}, \ and\ \bibinfo {author}
  {\bibfnamefont {J.~C.}\ \bibnamefont {Cuevas}},\ }\bibfield  {title} {\emph
  {\bibinfo {title} {Interplay between {Yu}-{Shiba}-{Rusinov} states and
  multiple {Andreev} reflections},}\ }\href {\doibase
  10.1103/PhysRevB.101.235445} {\bibfield  {journal} {\bibinfo  {journal}
  {Physical Review B}\ }\textbf {\bibinfo {volume} {101}},\ \bibinfo {pages}
  {235445} (\bibinfo {year} {2020})}\BibitemShut {NoStop}%
\bibitem [{\citenamefont {Huang}\ \emph
  {et~al.}(2020{\natexlab{b}})\citenamefont {Huang}, \citenamefont {Drost},
  \citenamefont {Senkpiel}, \citenamefont {Padurariu}, \citenamefont {Kubala},
  \citenamefont {Yeyati}, \citenamefont {Cuevas}, \citenamefont {Ankerhold},
  \citenamefont {Kern},\ and\ \citenamefont {Ast}}]{si_huang_quantum_2020}%
  \BibitemOpen
  \bibfield  {author} {\bibinfo {author} {\bibfnamefont {H.}~\bibnamefont
  {Huang}}, \bibinfo {author} {\bibfnamefont {R.}~\bibnamefont {Drost}},
  \bibinfo {author} {\bibfnamefont {J.}~\bibnamefont {Senkpiel}}, \bibinfo
  {author} {\bibfnamefont {C.}~\bibnamefont {Padurariu}}, \bibinfo {author}
  {\bibfnamefont {B.}~\bibnamefont {Kubala}}, \bibinfo {author} {\bibfnamefont
  {A.~L.}\ \bibnamefont {Yeyati}}, \bibinfo {author} {\bibfnamefont {J.~C.}\
  \bibnamefont {Cuevas}}, \bibinfo {author} {\bibfnamefont {J.}~\bibnamefont
  {Ankerhold}}, \bibinfo {author} {\bibfnamefont {K.}~\bibnamefont {Kern}}, \
  and\ \bibinfo {author} {\bibfnamefont {C.~R.}\ \bibnamefont {Ast}},\
  }\bibfield  {title} {{\selectlanguage {english}\emph {\bibinfo {title}
  {Quantum phase transitions and the role of impurity-substrate hybridization
  in {Yu}-{Shiba}-{Rusinov} states},}\ }}\href {\doibase
  10.1038/s42005-020-00469-0} {\bibfield  {journal} {\bibinfo  {journal}
  {Communications Physics}\ }\textbf {\bibinfo {volume} {3}},\ \bibinfo {pages}
  {199} (\bibinfo {year} {2020}{\natexlab{b}})}\BibitemShut {NoStop}%
\bibitem [{\citenamefont {J\"ack}\ \emph {et~al.}(2016)\citenamefont {J\"ack},
  \citenamefont {Eltschka}, \citenamefont {Assig}, \citenamefont {Etzkorn},
  \citenamefont {Ast},\ and\ \citenamefont {Kern}}]{si_jack_critical_2016}%
  \BibitemOpen
  \bibfield  {author} {\bibinfo {author} {\bibfnamefont {B.}~\bibnamefont
  {J\"ack}}, \bibinfo {author} {\bibfnamefont {M.}~\bibnamefont {Eltschka}},
  \bibinfo {author} {\bibfnamefont {M.}~\bibnamefont {Assig}}, \bibinfo
  {author} {\bibfnamefont {M.}~\bibnamefont {Etzkorn}}, \bibinfo {author}
  {\bibfnamefont {C.~R.}\ \bibnamefont {Ast}}, \ and\ \bibinfo {author}
  {\bibfnamefont {K.}~\bibnamefont {Kern}},\ }\bibfield  {title} {\emph
  {\bibinfo {title} {Critical {Josephson} current in the dynamical {Coulomb}
  blockade regime},}\ }\href {\doibase 10.1103/PhysRevB.93.020504} {\bibfield
  {journal} {\bibinfo  {journal} {Phys. Rev. B}\ }\textbf {\bibinfo {volume}
  {93}},\ \bibinfo {pages} {020504} (\bibinfo {year} {2016})}\BibitemShut
  {NoStop}%
\bibitem [{\citenamefont {Ambegaokar}\ and\ \citenamefont
  {Baratoff}(1963)}]{si_ambegaokar_tunneling_1963}%
  \BibitemOpen
  \bibfield  {author} {\bibinfo {author} {\bibfnamefont {V.}~\bibnamefont
  {Ambegaokar}}\ and\ \bibinfo {author} {\bibfnamefont {A.}~\bibnamefont
  {Baratoff}},\ }\bibfield  {title} {\emph {\bibinfo {title} {Tunneling between
  superconductors},}\ }\href {\doibase 10.1103/PhysRevLett.10.486} {\bibfield
  {journal} {\bibinfo  {journal} {Phys. Rev. Lett.}\ }\textbf {\bibinfo
  {volume} {10}},\ \bibinfo {pages} {486} (\bibinfo {year} {1963})}\BibitemShut
  {NoStop}%
\end{thebibliography}
\end{document}